\documentclass[12pt]{article}
\usepackage[top=1in,bottom=1in,left=1in,right=1in]{geometry}
\usepackage{graphicx} 
\usepackage{float}
\usepackage{booktabs}
\usepackage{adjustbox}
\usepackage{algorithm}
\usepackage{algorithmicx}
\usepackage{algpseudocode}
\usepackage{amsmath,amssymb, amsthm}
\usepackage{xcolor}
\usepackage{xr}
\usepackage{setspace}
\usepackage{chngcntr}
\usepackage{enumitem}
\usepackage{multirow}

\usepackage{hyperref}
\hypersetup{
    colorlinks=true,
    urlcolor=cyan
}
\usepackage[font=footnotesize]{caption}
\usepackage{natbib}
\makeatletter
\renewcommand{\ALG@name}{}
\makeatother

\theoremstyle{plain}
\newtheorem{definition}{Definition}
\newtheorem{proposition}{Proposition}

\newtheorem{theorem}{Theorem}
\newtheorem{assumption}{Assumption}

\usepackage{mdframed}
\newmdenv[
  topline=false,
  bottomline=false,
  rightline=false,
  skipabove=\topsep,
  skipbelow=\topsep,
  innertopmargin=0,
  linewidth=0.8pt
]{leftrule}

\title{Inspection-Guided Randomization: \\
A Flexible and Transparent Restricted Randomization Framework for Better Experimental Design}
\author{
Maggie Wang\thanks{Department of Biomedical Data Science, Stanford University}\and 
René F. Kizilcec\thanks{Department of Information Science, Cornell University}\and 
Michael Baiocchi\thanks{Department of Epidemiology and Population Health, Stanford University, Email: \href{mailto:baiocchi@stanford.edu}{baiocchi@stanford.edu}}
}
\date{January 29, 2025}

\begin{document}
\maketitle

\begin{abstract}
    Randomized experiments are considered the gold standard for estimating causal effects. However, out of the set of possible randomized assignments, some may be likely to produce poor effect estimates and misleading conclusions. Restricted randomization is an experimental design strategy that filters out undesirable treatment assignments, but its application has primarily been limited to ensuring covariate balance in two-arm studies where the target estimand is the average treatment effect. Other experimental settings with different design desiderata and target effect estimands could also stand to benefit from a restricted randomization approach. We introduce Inspection-Guided Randomization (IGR), a transparent and flexible framework for restricted randomization that filters out undesirable treatment assignments by inspecting assignments against analyst-specified, domain-informed design desiderata. In IGR, the acceptable treatment assignments are locked in ex ante and pre-registered in the trial protocol, thus safeguarding against \textit{p}-hacking and promoting reproducibility. Through illustrative simulation studies motivated by behavioral health and education interventions, we demonstrate how IGR can be used to improve effect estimates compared to benchmark designs in experiments with interference and in group formation experiments.
    \end{abstract}
    
\section{Introduction}\label{sec:intro}

Randomized experiments are considered the gold standard for estimating causal effects of interventions. By randomizing the treatment, the distribution of baseline covariates is similar in expectation across treatment arms, and observed differences in the outcome can be attributed to the intervention. However, randomization alone does not guarantee covariate balance for every realized allocation of treatments \citep{Fisher1935-mh, Rubin2008-va}. Suppose the treatment allocation chosen for an experiment, by chance, exhibits imbalance on covariates. At best, the analyst makes careful \textit{ex post} adjustments for imbalance, usually according to a regression model that relies on assumptions that may or may not hold \citep{Freedman2008-bf}. At worst, covariate imbalance is overlooked or handled improperly (e.g., in an attempt to \textit{p}-hack), leading to unreliable conclusions about treatment effects (e.g., false positives). Restricted randomization helps prevent imbalanced treatment allocations. In restricted randomization, the assignment mechanism is modified such that the probability of choosing an undesirable allocation is set to zero. While performing restricted randomization is rather common, it is primarily applied to the canonical use case of balancing two-arm trials where the target effect estimand is the average treatment effect. 

Experimental settings with different design desiderata and estimands could also stand to benefit from a restricted randomization design. A natural extension of balancing a two-arm trial is to balance a multi-arm trial, where the goal is to simultaneously estimate treatment effects for multiple interventions. Additionally, in experiments where we want to explore treatment effect moderators \citep{VanderWeele2015-iq}, we may want to construct treatment arms with intentional ``imbalance" on a hypothesized moderating covariate but maintain balance on other covariates. We may also be interested in controlling factors beyond balance that could jeopardize the credibility of effect estimates. Causal inference in the presence of interference \citep{Cox1958-xg, Rosenbaum2007-xe}, when one individual's treatment affects a different individual's outcome, has been an area of growing interest \citep{Hudgens2008-os, aronow2017-gb, Eckles2017-sr, Athey2018-yf}. For many target effect estimands, interference should be minimized to obtain unbiased effect estimates. Here, if we have some \textit{a priori} knowledge of how interference is likely to occur, we can use restricted randomization to rule out allocations that are likely to generate large amounts of interference. 

More broadly, the core task in experimental design is to set up the study such that it generates a trustworthy answer to the study's motivating causal question. We introduce \textbf{Inspection-Guided Randomization} (IGR), a flexible and transparent framework for restricted randomization in which the study analyst \textit{inspects} the desirability of candidate randomization allocations, then constrains the candidate space to produce less biased and more precise estimates of the target estimand. The flexibility in IGR exists along two fronts. First, IGR permits targeting diverse design desiderata beyond those that relate to balance, thus expanding the scope of use cases for restricted randomization. Second, IGR incorporates a checkpoint that encourages the study analyst to carefully evaluate, and potentially adapt, the construction of the restricted assignment mechanism. IGR promotes transparency by enforcing that the set of acceptable treatment allocations be pre-registered in the trial protocol prior to running the study. 

This work is organized as follows. In Section \ref{sec:rel}, we review prior work in restricted randomization and experimental design, then highlight the contributions we make through IGR. In Section \ref{sec:igr}, we introduce notation and an overview of the IGR framework. In Section \ref{sec:igr-detail}, we walk through each step of IGR in detail and use a simulation to illustrate how it can be applied to the design of an experiment with interference. In Sections \ref{sec:igr-analysis}-\ref{sec:igr-effect-estimates}, we discuss analysis and inference for IGR-designed experiments and show how IGR can improve effect estimates. We conclude in Section \ref{sec:disc} with a summary of the benefits of using IGR and directions for future work.
    
\section{Related Work}\label{sec:rel}

\paragraph{Blocking and Matching} Blocking is a form of restricted randomization that rules out allocations that are imbalanced on one or more blocked covariates and has been used widely in the design of experiments \citep{Fisher1992-ed, Box2005-we}. Matched pair designs use matching algorithms to pair experimental units on the basis of a metric of covariate similarity, thus facilitating balance on multiple categorical and continuous covariates at once \citep{Greevy2004-pd, Xu2010-ha, Imai2009-fq}. More recently, efficient blocking algorithms have been developed that extend matched-pair designs and allow for balanced blocks of flexible sizes \citep{Higgins2016-tm}.

\paragraph{Tightening, Constraining, and Re-randomizing} An alternative approach to blocking and matching involves first generating a candidate set of allocations according to a pre-specified mechanism, then selecting one allocation from among this set as the official randomization to deploy. Tukey referred to this procedure as ``tightening" the clinical trial \citep{Tukey1993-no}, and Moulton later formalized a ``covariate-constrained randomization" design for cluster-randomized trials \citep{Moulton2004-sj}. There has also been work in using constrained randomization in multi-arm settings with the use of multi-arm balance metrics \citep{Ciolino2019-ms, Watson2021-ob, Zhou2021-rq}. With constrained designs, inference can be done via a randomization test over the candidate allocations.

Instead of listing all possible allocations at once, re-randomization repeatedly randomizes until producing the first allocation that meets pre-specified criteria \citep{Rubin2008-va, Cox2009-cy, Morgan2012-mb}. Re-randomization has been used beyond two-arm study designs in experiments with factorial designs \citep{Branson2016-fm} and in experiments on networks where units have correlated potential outcomes \citep{Basse2018-ii}. Although re-randomization can save computing power, it opens the door to \textit{p}-hacking because the set of criteria-meeting allocations used to conduct a randomization test are sampled after study outcomes have been observed (see further discussion in Appendix \ref{app-sec:phack-rerand} on the risk of Type I error rate inflation when the accepted allocations in the analysis phase are different from those in the design phase). Additionally, re-randomization assumes that the analyst's first choice of restriction criteria produces a satisfactory design, when in reality the criteria may permit allocations that have poor properties for estimation and inference. 

\paragraph{Optimal and Model-assisted Design} \cite{Kasy2016-wo} argues that if the goal of the experiment is to minimize ``disparity" between the estimated and true treatment effect (e.g., mean-squared error), one should deploy the single allocation that optimizes the expected loss over possible data-generating processes. Identifying an optimal allocation in this way, however, requires specifying a prior distribution for the loss function. Model-assisted design is similarly motivated, but optimizes the expected mean-squared error over random data samples from a specific data-generating process rather than for a distribution of data-generating processes \citep{Basse2018-ii}. Unlike with constrained randomization and re-randomization, when only a single optimal allocation is selected, randomization inference is no longer warranted. 

\paragraph{Reproducible Science} Results of published studies are often not reproducible because they were generated through the use of questionable research practices aimed at ``publication-worthiness" rather than scientific correctness and robustness \citep{Baker2016-hy, Munafo2017-tb, Ioannidis2005-dx}. These practices include \textit{p}-hacking, reworking research hypotheses after the results of a study have already been collected, and over-interpreting the data \citep{Munafo2017-tb, Ioannidis2005-dx}. To promote reproducible and open science, there has been a push for pre-registering study protocols that hold researchers accountable to a planned study design and set of analyses \citep{Munafo2017-tb}.

\paragraph{Our contributions} 
IGR unifies and builds upon ideas across several of the works above. Tightening, re-randomization, and covariate-constrained randomization were developed to target balance criteria only, and subsequent literature that applies or expands these randomization strategies also only consider balance (Appendix Table \ref{tab:rerand-lit-review}). While IGR resembles covariate-constrained randomization, we make an important extension by recognizing that it can be beneficial to inspect allocations on the basis of multiple properties, particularly properties other than balance. IGR is also related to model-assisted design \citep{Basse2018-ii}, but whereas model-assisted design requires deriving a closed-form expression for mean-squared error, IGR can use domain-informed, approximate metrics when the expression for mean-squared error is unknown or difficult to derive. 

Additionally, most prior restricted randomization strategies, e.g. re-randomization, use a naive sampling approach to select an allocation. In contrast, in IGR, we can efficiently ``search" through a large candidate allocation space (e.g., with genetic algorithms) to enumerate a set of candidate allocations that score well according to specified metrics. Speaking more generally, in re-randomization, the process of searching for an allocation and the process of randomizing are conflated in the act of sampling. IGR, on the other hand, separates the enumeration step, which can be done without any randomization, from the randomization step, and thus draws a clear distinction between searching and randomizing. Some implementations of constrained randomization use a similar approach to IGR \citep{Moulton2004-sj, De_Hoop2012-bm, Li2016-zf, Ciolino2019-ms, Watson2021-ob, Zhou2021-rq}, where a candidate set of allocations are first enumerated, and the final allocation is chosen randomly from among criterion-meeting allocations within the candidate set. However, these implementations use a random sampling procedure to produce the initial candidate set; they do not highlight that enumerating the candidate set can involve a directed search and does not necessarily need to be random. The randomness used for inference occurs later, in the actual assignment.

In line with efforts to make studies more reproducible, IGR includes and enforces a step where accepted allocations are pre-registered. As long as the study analyst performs randomization inference over the pre-registered set of accepted allocations, then getting the nominal Type I error rate is guaranteed. Pre-registration is not explicitly present in any existing restricted randomization framework, so study analysts are not held accountable to conducting correct analyses. We show, for instance, that in re-randomization, we can easily inflate the Type I error rate by using a more restrictive balance criterion in the analysis phase than in the design phase. 

Finally, IGR places an emphasis on deliberately interrogating the quality of the restricted design and on making subsequent adaptations to the restriction as necessary. This adaptation step is either under-emphasized or absent from prior approaches to restricted randomization. In practice, we suspect that two common forms of adaptation will be: (i) fixing inspection metrics that do not correctly target the design desiderata (e.g., a balance metric that does not capture the kind of imbalance concerning to the study stakeholders) and (ii) adjusting the relative emphases placed on different design desiderata to better navigate trade-offs (e.g., in trials with interference, homophily can create tension between achieving covariate balance and minimizing interference, and which of the two desiderata is prioritized may need to be adjusted).

Several prior real-world studies have been designed with restricted randomization techniques related to IGR that included some, but not all, of the elements of the IGR framework \citep{Murnane2023-vy, Cho2021-pf, Cho2022-oy, Kizilcec2022-ul, Zahrt2023-jd}. These studies demonstrate the need for a formal framework like IGR that unifies design principles, and they also highlight the potential for IGR to be applicable and useful in practice. In this work, we present simulated vignettes of an experiment with interference and of a group formation experiment (Section \ref{sec:igr-detail} and Appendix \ref{app-sec:group-form}, respectively) to explain specific features of the IGR framework and to highlight how IGR can be used for diverse experimental settings, including and beyond those where related restricted randomization techniques have been used in the past.

\section{An Overview of Inspection-Guided Randomization}\label{sec:igr}
\subsection{Setup and notation}\label{sec:igr-setup}
We consider a setting with $N$ participants in a $k$-armed study, where the participants are indexed by $i = 1,...,N$. Let $\mathbf{X}$ be the $N$-by-$p$ dimensional matrix representing the participants’ baseline covariates.  For a given participant $i$, we use the notation $Z_i$ to denote the participant’s treatment assignment, where $Z_i \in \{0,...,k-1\}$. Let the $N$-dimensional vector of treatments be denoted $\mathbf{Z} = (Z_1, Z_2, \dots, Z_N)$. Since treatment is assigned at random, $Z_i$ is a random variable, and it follows that $\mathbf{Z}$ is a random vector. We denote a particular realization of a treatment assignment for individual $i$ as lowercase $z_i$ and likewise denote the intervention assignment vector lowercase $\mathbf{z}$. The probability that random vector $\mathbf{Z}$ takes on a particular value $\mathbf{z}$ is written $P(\mathbf{Z} = \mathbf{z}) = p_{\mathbf{z}}$, and we refer to $P(\mathbf{Z})$ as the assignment mechanism. We call a particular realization of the intervention assignment vector a treatment allocation. Letting $\mathcal{Z}$ denote the collection of all $k^N$ possible values of $\mathbf{z}$, we have that $\sum_{\mathbf{z} \in \mathcal{Z}} p_{\mathbf{z}} = 1$. 

\subsection{Outline of the IGR framework}\label{sec:methods-igr}
For clarity and ease of use in practice, we organize the IGR design framework into distinct steps. We describe each in brief below, then elaborate further in the next section:
\begin{enumerate}[itemsep=-1ex]
    \item \textbf{Specification}. The analyst specifies the target effect estimand and the fitness function used to measure the desirability of each candidate treatment allocation. 
    \item \textbf{Enumeration}. The analyst enumerates a large pool of candidate allocations using a simple assignment mechanism. 
    \item \textbf{Restriction}. The analyst scores each candidate allocation with the specified fitness function and filters out those that are low-scoring, according to a restriction rule.
    \item \textbf{Evaluation \& Adaptation}. The analyst evaluates their choice of fitness function and restriction rule, then makes adaptations if needed to better achieve design desiderata.
    \item \textbf{Pre-registration}. The analyst pre-registers the IGR assignment mechanism.  
    \item \textbf{Randomization}. The analyst samples one allocation to deploy for the experiment.
\end{enumerate}

\section{IGR in Detail, with Application to a Simulated Experiment With Interference}\label{sec:igr-detail}
In this section, we provide a full description and rationale for the steps of the IGR framework. To make the steps more concrete, we also illustrate how to apply them to a simulated experiment with interference. We begin this section by describing the domain and set-up for the simulated experiment, then walk through the steps of IGR in detail, explaining how each can be applied in the context of the simulated experiment. Note that IGR can be used in many experimental settings, not just those with interference. In Appendix \ref{app-sec:group-form}, for example, we also highlight how IGR can be applied to a group formation experiment. All simulation code is available at \url{https://github.com/wangmagg/inspection-guided-randomization}.

\subsection{Simulated Experiment with Interference}\label{subsec:sim-exp-interf}

\paragraph {The challenge of interference}
Interference occurs when the treatment assigned to one observational unit affects the outcome of a different unit. Interference is often present with behavioral interventions, where information can diffuse or spill over along social ties \citep{Valente2012-fj, Cai2015-kn, Kim2015-dk}. When interference is present, a common target treatment effect estimand is the global average treatment effect, or a comparison of the average outcome if everyone were to receive treatment versus if everyone were to receive control \citep{Hudgens2008-os}. Failing to properly account for interference can make estimates of the global average treatment effect both biased and imprecise \citep{Eckles2017-sr}. This in turn can result in misleading conclusions (e.g., false nulls) and misinformed recommendations for policy-making (e.g., failure to implement an effective intervention).

A common approach to controlling interference is to use cluster randomization, where randomization is done on the cluster level such that all individuals in the same cluster receive the same treatment \citep{Hayes2017-bn}. Given enough separation between clusters, there is likely no spillover between different treatment arms, and comparing treated to control clusters can give an unbiased estimate of the global average treatment effect. However, it may sometimes be difficult to identify clusters that are sufficiently separated to avoid any inter-cluster interference. In other cases, it may be possible to identify a few large clusters that are well-separated, such as villages that are far apart geographically. However, randomizing only a few clusters leads to low power, especially if there is strong intra-cluster correlation on prognostic covariates. Often, for clusters that form organically, individuals within a cluster are more similar than those in different clusters, so intra-cluster correlation is expected.

Within the IGR design framework, we can be more flexible with the way that we control interference. Leveraging prior knowledge, which may not necessarily be complete or fully correct, we can use an interference-targeting constraint to control the amount of anticipated interference in a more precise way than with cluster randomization. Even if we do not have perfect knowledge of the interference structure (e.g., which individuals interact to produce spillover), we can still develop ``coarse" metrics that measure interference. 

\paragraph{Simulation set-up}
Our simulated experiment with interference is based on a trial of a sexual assault prevention intervention for adolescent girls conducted across schools in urban settlements in Kenya. During the intervention, the girls learned and practiced verbal and physical skills for resisting sexual assault \citep{Baiocchi2017-om, Sarnquist2023-ho}.

An option for the design of this trial was to randomize the intervention at the settlement level, which would have strongly mitigated interference. However, the settlements were known to have a wide range of baseline rates of sexual assault, meaning randomizing on the settlement level would have made it difficult to achieve balance. Hence, the trial was instead randomized on the school level, and, because the settlements were densely populated, it is likely that girls attending different schools interacted with one another. The skills learned by girls assigned to receive the intervention may have been passed on to girls assigned to the control arm, resulting in interference.

In our simulation, we generate synthetic schools in five Kenyan settlements: Kibera, Mukuru, Huruma, Korogocho, and Dandora. Baseline covariates are sampled using a nested hierarchical model so that individuals attending the same schools are more similar than individuals attending different schools, and individuals within the same settlement are more similar than individuals across different settlements. We use a statistical network model to generate the interference network, where the probability two individuals are connected by an edge is inversely proportional to the $\gamma^{th}$ power of the Euclidean distance between the schools. To generate the observed outcome, we apply an additive treatment effect and assume that an individual is exposed to treatment if they directly receive treatment or if a certain fraction of their network neighbors receive treatment. Further details on the data-generating process and the parameter settings used in the simulations can be found in Appendix \ref{app-sec:interference-dgp}. 

\subsection{Step-by-Step Application of IGR}\label{subsec:igr-step-by-step}

We now go through IGR step-by-step in detail. After each step is described, we show how it can be applied to designing the simulated experiment.

\subsubsection{Specification}
In the \textit{Specification} step, the study analyst first works with the domain experts on the research team to specify the goals of the study in terms of the causal question of interest and the target effect estimand. Based on these goals, the team compiles desiderata for the study such that if the deployed randomization allocation were guaranteed to satisfy all desiderata, then the data collected would likely be conducive to credible causal inference.

The analyst then constructs a set of inspection metrics that are used to assess candidate allocations, where each metric measures the extent to which the allocation satisfies a particular desideratum (e.g., standardized mean difference as a metric to measure balance). In Table \ref{tab:inspection-metrics}, we list three broad categories of common design desiderata and examples of inspection metrics that might be used within each category.

\begin{table}[ht!]
    \centering
    \begin{tabular}{p{0.3 
    \linewidth}p{0.3\linewidth}p{0.33 \linewidth}}
    \toprule
         Desideratum Category &  Example Desideratum & Example Inspection Metric(s) \\
    \midrule  
         \textbf{Estimation properties} & Low variance & Standardized mean difference, Mahalanobis distance \\
         & Low bias in the presence of network or spatial spillover & Connectivity in a graph network, geographic proximity between treated/control units \\
         \textbf{Study logistics} & Ease of delivering a treatment for an infectious disease & Travel time or distance to reach all units assigned to the treatment arm \\
         \textbf{Conditions for testing scientific theory} & Placing individuals in certain social contexts to study effect on academic achievement & Compositional characteristics of the individuals assigned to a given classroom (e.g. gender ratio) \\
    \bottomrule
    \end{tabular}
    \caption{Common desiderata categories, specific examples of desiderata within each category, and example inspection metrics that can be used to measure the extent to which an allocation satisfies a particular desideratum.}
    \label{tab:inspection-metrics}
\end{table}

The analyst also decides on an aggregator function that combines the separate metric scores and summarizes an allocation's desirability. In certain settings, the desiderata may be conflicting, where scoring better for some desiderata necessarily entails worse scores on other desiderata. The aggregator can be used to navigate such desiderata trade-offs, for instance by incorporating a weighting scheme that prioritizes certain desiderata over others. Let $f$ be the fitness function that maps an allocation $\mathbf{z}$ to its score by combining the outputs of each of the individual inspection metrics via the aggregation function.

Finally, the analyst specifies a restriction rule, $r$, that constrains the space of candidate treatment allocations on the basis of each allocation's score, where $r(f(\mathbf{z}, \mathbf{X})) = 0$ means $\mathbf{z}$ is filtered out by setting $p_{\mathbf{z}}$ to zero and $r(f(\mathbf{z}, \mathbf{X})) = 1$ means $\mathbf{z}$ is deemed acceptable by setting $p_{\mathbf{z}}$ to a positive value between 0 and 1. For example, $r$ could be a thresholding rule, where only allocations with scores below some threshold $s^*$ are accepted.

\begin{leftrule}

\paragraph{\textit{Specification} in a Simulated Experiment}
We specify the target effect estimand to be the global average treatment effect, $\tau_{GATE} = \frac{1}{N}\sum_{i=1}^N Y_i(\mathbf{1}) - Y_i(\mathbf{0})$ where $\mathbf{1}$ is the all-ones vector (everyone is treated) and $\mathbf{0}$ is the all-zeros vector (everyone is control). Our design goal is to assign treatments such that those in the treatment (control) arm experience a universe that is as close as possible to an all-treated (all-control) universe while also balancing covariates across arms. Thus, we specify two inspection metrics, one to measure interference and one to measure balance.

For interventions that spread via person-to-person interaction, the likelihood of such spread or spillover may decrease with geographic distance between individuals. While we may not know precisely which individual(s) will interact with which other individual(s), we can try to allocate treatments in such a way that no two individuals assigned to treatment and control are too close to one another geographically. Closeness can be defined, for example, in terms of where they live, where they attend school, and where they work. This line of reasoning motivates an inspection metric that measures the minimum distance between any two individuals assigned to treatment and control, where a larger minimum distance is more desirable. To stay consistent with smaller scores being better, we use the reciprocal of the minimum distance. Formally, we define inspection metric $m^{(\texttt{InvMinEuclideanDist})}$ as
    \begin{align}\label{eq:metric-euclid}
        m^{(\texttt{InvMinEuclideanDist})} = 
            \left ( 
            \min_{(i, j) \in 
            [N] \times [N], z_i \neq z_j} ||\mathbf{l}_i - \mathbf{l}_j||_2^2
            \right )^{-1}
    \end{align}
where $z_i, z_j \in \{0, 1\}$ are the treatment assignments for units $i$ and $j$, respectively and $\mathbf{l}_i, \mathbf{l}_j \in \mathbb{R}^2$ are coordinates for $i$ and $j$. While we use Euclidean distance here, other distance functions can also be used. It is also possible to expand upon a simple distance function by incorporating known information about ease of travel, such as the presence or absence of interconnecting roads. For balance, we use Mahalanobis distance,
\begin{align}
     m^{(\texttt{Mahalanobis})}(\mathbf{z}, \mathbf{g}, \mathbf{X}) &= 
        \left( \bar{\mathbf{X}}^{(1)} - \bar{\mathbf{X}}^{(0)} \right)
        \mathbf{S}^{-1}
        \left( \bar{\mathbf{X}}^{(1)} - \bar{\mathbf{X}}^{(0)} \right)^T
\end{align}

where $\bar{\mathbf{X}}^{(1)}$ and $\bar{\mathbf{X}}^{(0)}$ are the mean covariate vectors in the treated and control arms, respectively, and $\mathbf{S}$ is the sample covariance matrix. Combining the interference and balance metrics, we define the fitness function

\begin{equation}\label{eq:euclid+mahalanobis}
\begin{split}
    f^{(\texttt{I+M})}(\mathbf{z}, \mathbf{X})
    &= 0.5 \cdot m^{(\texttt{InvMinEuclideanDist})}(\mathbf{z}, \mathbf{X}) + 0.5 \cdot m^{(\texttt{Mahalanobis})}\cdot(\mathbf{z}, \mathbf{X}) 
\end{split}
\end{equation}

To put the metrics on the same scale, we standardize the values across all enumerated candidate allocations (see the \textit{Enumeration} step below) before combining the metrics in the fitness function.

\end{leftrule}

\subsubsection{Enumeration}
In the \textit{Enumeration} step, the analyst enumerates a pool of unique candidate allocations, $\mathcal{Z}_{pool} = \{\mathbf{z_1}, \mathbf{z_2}, \dots \mathbf{z_M}\}$. For a $k$-armed study, there are $k^N$ total possible allocations. When $N$ is large, it may be necessary to set $M << k^N$. The enumeration of $\mathcal{Z}_{pool}$ can be done by sampling from any choice of assignment mechanism $P$, including Bernoulli randomization and complete randomization. In some cases, sampling from an assignment mechanism that is already restricted, such as a block randomization, can increase the efficiency of the enumeration step, since the mechanism inherently rules out some of the undesirable allocations (see Appendix \ref{app-sec:group-form} for an example). 

Since $M << k^N$, it is possible that many desirable, high-scoring allocations are overlooked and not enumerated as part of $\mathcal{Z}_{pool}$. Instead of using simple enumeration strategies, we may instead want to efficiently search through the space of all $k^N$ possible allocations to construct a more optimal $\mathcal{Z}_{pool}^{(*)}$. One approach to construct $\mathcal{Z}^{(*)}_{pool}$ is to use genetic algorithms \citep{Mitchell1998-sg}. The initial $\mathcal{Z}_{pool}$ is viewed as a population of treatment allocations. These allocations ``mate" in pairs to produce a new generation of allocations, where novel allocations are introduced through ``mutation" (single entries of an allocation are changed) and ``crossover" (segments of two parent allocations are swapped). The individuals in this generation are scored using the specified fitness function $f$, and those with the best scores survive and breed to produce the next generation. We can repeat this process over $T$ generations to obtain $\mathcal{Z}_{pool}^{(*)}$. 

\begin{leftrule}
\paragraph{\textit{Enumeration} in a Simulated Experiment} We first sample candidate allocations from a complete randomization assignment mechanism that assigns an equal number of schools to the treatment and control arms. We enumerate a total of $M = 100,000$ allocations in the initial candidate pool, which is in line with prior work \citep{Li2016-zf, Watson2021-ob, Nordin2022-jw}. We then apply 3 generations of mutations and crossover to the initially enumerated pool to search for a pool with better fitness scores.
\end{leftrule}

\subsubsection{Restriction}
The analyst selects a restriction rule $r$, then applies the fitness function and restriction rule to the enumerated pool of candidate allocations to produce the following assignment mechanism:
    \begin{align*}
        P_{IGR}(\mathbf{Z} = \mathbf{z}) = 
        \begin{cases}
        0 & \mathbf{z} \notin \mathcal{Z}_{pool} \\
        0 & \mathbf{z} \in \mathcal{Z}_{pool}, (r \circ f)(\mathbf{z}, \mathbf{X}) = 0 \\
        p_{\mathbf{z}} & \mathbf{z} \in \mathcal{Z}_{pool}, (r \circ f)(\mathbf{z}, \mathbf{X}) = 1
        \end{cases}
    \end{align*}
We refer to allocations that have a non-zero probability of being sampled as ``accepted" allocations. At this step, the assignment mechanism $P_{IGR}$ is proposed but not yet locked in. 

\begin{leftrule}
\paragraph{\textit{Restriction} in a Simulated Experiment} We use the restriction rule $r(s) = 0 \text{ if } \ s \geq s^*, r(s) = 1 \text{ if } s < s^* $, where $s^*$ is the $0.5^{\text{th}}$-percentile of scores for the enumerated candidate allocations. This restriction rule filters out all but the top $0.5^{\text{th}}$-percentile of the $100,000$ enumerated allocations, resulting in $m = 500$ accepted allocations.
\end{leftrule}

\subsubsection{Evaluation \& Adaptation}
Before proceeding and locking in the restricted assignment mechanism, the study analyst and domain experts pause to evaluate the fitness function $f$ and restriction rule $r$, making adaptations and improvements to them if necessary. Even if $f$ and $r$ were carefully identified in \textit{Specification}, they still may be flawed in unforeseen ways and could yield poor study designs. We highlight three dimensions along which $f$ and $r$ should be evaluated: discriminatory power,  desiderata trade-offs, and over-restriction. 
\begin{enumerate}
    \item Discriminatory power: Fitness functions should provide meaningful differentiation between candidate allocations. If the fitness function used to score allocations produces a similar score for all candidates, then there would be no reason to expect that using a restricted assignment mechanism would confer any improvements in effect estimation. When confronted with a homogeneous set of scores, the fitness function may require redefinition, for instance through introducing additional inspection metrics.
    
    \item Desiderata trade-offs: When there are multiple competing design desiderata, it may not be immediately clear whether and how to best construct the fitness function to weigh desiderata trade-offs. Visualizing the space of all candidate allocations by plotting the score outputs of different inspection metrics on separate axes can assist in understanding the nature of desiderata trade-offs. Then, visualizing the space of accepted allocations can assist in understanding whether the fitness function navigates such trade-offs as desired. If necessary, the analyst can make informed adjustments to the fitness function by modifying the aggregator function that combines the inspection metric scores.
    
    \item Over-restriction: We consider two types of over-restriction. In the first type of over-restriction, the combination of fitness function and restriction rule may result in too few accepted allocations to obtain \textit{p}-values with the desired level of granularity (see Section \ref{sec:igr-analysis} for details on deriving \textit{p}-values from randomization tests). When this occurs, the analyst may need to modify the restriction rule to be less aggressive or choose to apply a restriction rule that always selects the top $m$ allocations. Too few accepted allocations could also be a consequence of using an enumeration strategy that does not ``hit" the correct regions of the space of all possible allocations. This issue can be addressed by increasing the size of the enumerated pool or by using stochastic optimization routines. In the second type of over-restriction, the set of accepted allocations may be too similar to one another, such that multiple units or groups of units are assigned to the same treatment arm across most accepted allocations. If we rely on an asymptotic p-value in testing the null hypothesis, then this kind of over-restriction can result in departures from the nominal Type I error rate \citep{Moulton2004-sj}. An exact p-value obtained through randomization inference circumvents this issue, but highly correlated allocations can also lead to increased mean-squared error and reduced power, regardless of whether randomization inference is used \citep{Nordin2022-jw, Krieger2020-hl}. Relaxing the restriction rule $r$ to permit additional allocations is one way to alleviate over-restriction issues.
\end{enumerate}

\begin{figure}[!ht]
    \centering
    \includegraphics[width=\linewidth]{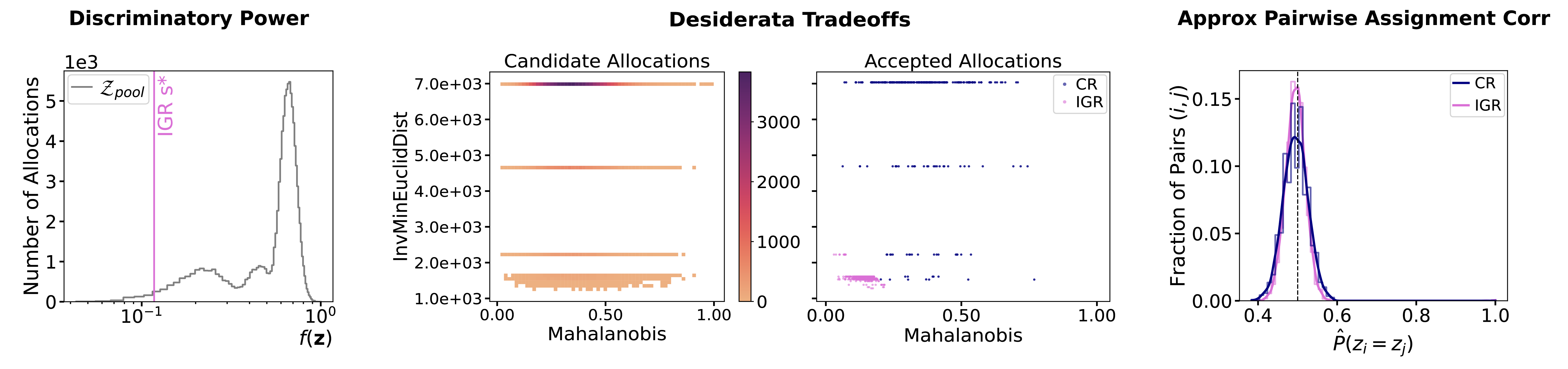}
    \caption{Visual checks for discriminatory power, desiderata trade-offs, and over-restriction (via approximate pairwise assignment correlation) under fitness function $f^{(\texttt{I+M})}$.}
    \label{fig:eval-adapt-iter1}
\end{figure}

\begin{leftrule}
\paragraph{\textit{Evaluation \& Adaption} in a Simulated Experiment}
We use visualizations to assist in evaluating our choice of fitness function and restriction rule. Additional toy examples and a more detailed discussion on the choice of visualizations can be found in Appendix \ref{app-sec:eval-adapt-elab}. Below, we show an example iterative process of evaluation and adaptation for our simulated experiment.

\textit{Iteration 1:}
To assess discriminatory power, we plot a histogram of the scores for the candidate allocations and check that the histogram has sufficient spread (Figure \ref{fig:eval-adapt-iter1}, left). A ``spiky" histogram where all fitness scores are concentrated around a few values is indicative of a fitness function that has low discriminatory power.  

To examine desiderata trade-offs in the pool of candidate allocations, we plot a two-dimensional histogram with the balance metric on the x-axis and the interference metric on the y-axis (Figure \ref{fig:eval-adapt-iter1}, middle). By looking at this heatmap, we can check whether allocations with better values for one metric tend to have worse values in the other metric. To examine desiderata trade-offs in the accepted set of candidate allocations, we make a scatterplot with the balance metric on the x-axis and the interference metric on the y-axis. Each point in the scatterplot corresponds to a single accepted allocation. By comparing the positioning of the accepted allocations to those of a benchmark randomization strategy, we can evaluate whether our restricted randomization scheme is navigating desiderata trade-offs as desired. 

To check for over-restriction, we plot a histogram of the approximate pairwise assignment correlation (Figure \ref{fig:eval-adapt-iter1}, right). Specifically, for each pair of units, $i, j$, we compute the fraction of candidate allocations where $i$ and $j$ are assigned to the same arm. When there is no pairwise assignment correlation, then each pair $i, j$ should appear in the same arm in half of the candidate allocations.
\end{leftrule}

\begin{figure}[!ht]
    \centering
    \includegraphics[width=\linewidth]{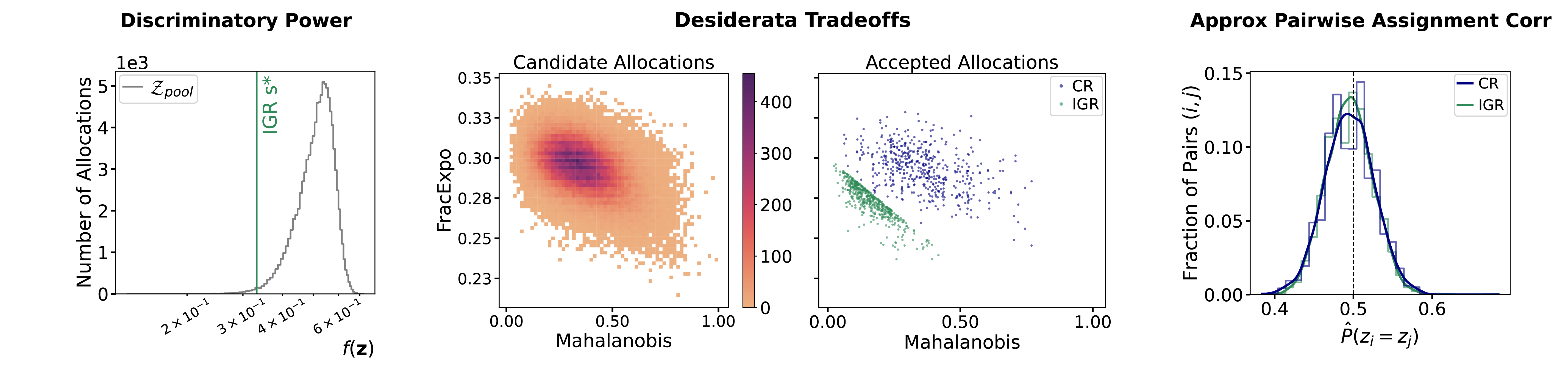}
    \caption{Visual checks for discriminatory power, desiderata trade-offs, and over-restriction (via approximate pairwise assignment correlation) under fitness function $f^{(\texttt{F+M})}$.}
    \label{fig:eval-adapt-iter2}
\end{figure}

\begin{leftrule}

\textit{Iteration 2:}
Based on the score histogram in Figure \ref{fig:eval-adapt-iter1}, it appears that the fitness function provides sufficient differentiation between allocations. Based on the desiderata trade-off plots, however, we can see that the interference metric is extremely coarse and takes on only a handful of distinct values. At the smallest value of the interference metrics, fine-grained fitness score differentiation is driven primarily by the balance metric. This may be appropriate if we are only interested in filtering out the most interference-prone allocations. If, on the other hand, we want to be more selective against interference even in the low-interference regimes, then a different interference metric is necessary. 

Suppose we are able to collect more detailed knowledge of a network of interactions between individuals. For instance, we can ask participants to list the friends with whom they communicate regularly and build a network based on each participant's list of contacts \citep{Cai2015-kn, Kim2015-dk}. In the network interference literature, it is common to assume that an individual is ``exposed" through the treatments of their neighbors in the network via a neighborhood exposure model \citep{aronow2017-gb, Manski2013-ol}. We therefore propose a new interference metric that counts the fraction of control individuals exposed to treatment through their network neighbors,
 \begin{align}\label{eq:metric-fracexpo}
        m^{(\texttt{FracExpo})}(\mathbf{z}, \mathbf{A}) = 
        \frac
        {\sum_{i=1}^N \mathbf{1}\{\sum_{j \in \mathcal{N}(i)} z_j > q \cdot |\mathcal{N}(i)|\} \cdot (1-z_i)}
        {\sum_{i=1}^N (1-z_i)}
    \end{align}
where $\mathbf{A}$ is an $N \times N$ adjacency matrix, $\mathcal{N}(i)$ is the set of nodes in the network that are connected to $i$ by an edge, $\mathbf{z}_{\mathcal{N}(i)}$ is the vector of treatment assignments for the individuals in this neighborhood, and $q$ is a scalar between 0 and 1 that defines an exposure threshold. In our simulation, we set $q$ to 0.25, so that a control individual is considered exposed if at least 25$\%$ of their neighbors are treated. The new fitness function is,
\begin{equation}\label{eq:fracexpo-eq-weight}
\begin{split}
    f^{(\texttt{F+M})}(\mathbf{z}, \mathbf{X})
    &= 0.5 \cdot m^{(\texttt{FracExpo})}(\mathbf{z}, \mathbf{X}) + 0.5 \cdot m^{(\texttt{Mahalanobis})}(\mathbf{z}, \mathbf{X}) 
\end{split}
\end{equation}
We re-make each of the plots from iteration 1 (Figure \ref{fig:eval-adapt-iter2}). From the desiderata trade-off plots we see that, as desired, the new interference metric based on network exposure is more granular than the metric based on distance. We also see that the new overall fitness function has adequate discriminatory power and has similar approximate pairwise assignment correlation to complete randomization.
\end{leftrule}

\begin{figure}[!ht]
    \centering
    \includegraphics[width=\linewidth]{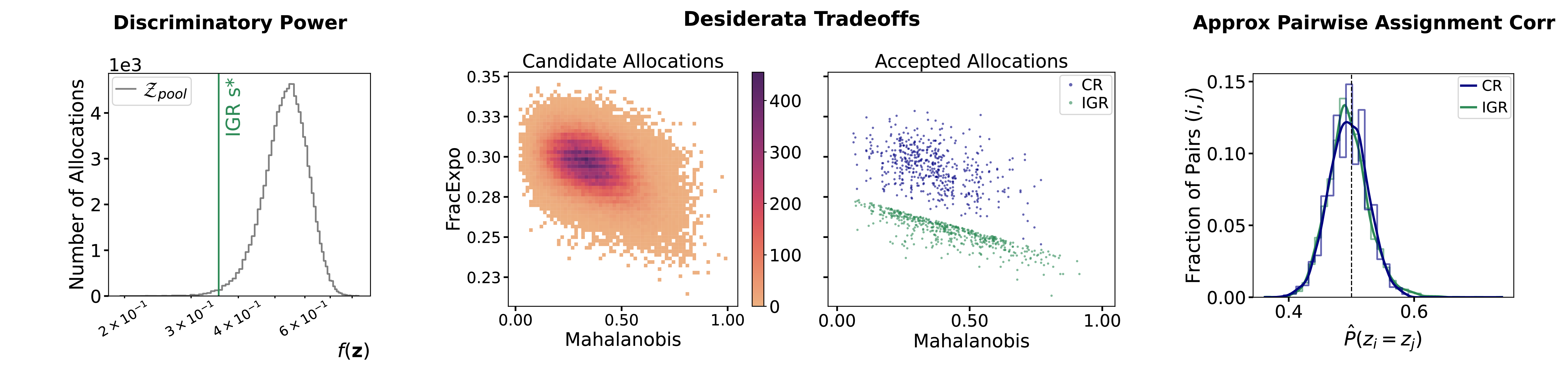}
    \caption{Visual checks for discriminatory power, desiderata trade-offs, and over-restriction (via approximate pairwise assignment correlation) under fitness function $f^{(\texttt{F+M})}$ with an upweighted interference metric.}
    \label{fig:eval-adapt-iter3}
\end{figure}

\begin{leftrule}
\textit{Iteration 3:}
We decide to use the interference metric based on network exposure. In previous iterations, the interference metric and balance metric were weighted equally. Now, we want to examine an alternative weighting scheme where more weight is applied to the interference metric. We therefore propose the fitness function,
\begin{equation}\label{eq:fracexpo-upweight-interf}
\begin{split}
    f^{(\texttt{F+M})}(\mathbf{z}, \mathbf{X})
    &= 0.75 \cdot m^{(\texttt{FracExpo})}(\mathbf{z}, \mathbf{X}) + 0.25 \cdot m^{(\texttt{Mahalanobis})}(\mathbf{z}, \mathbf{X}) 
\end{split}
\end{equation}
Again, we construct plots to check for discriminatory power, desiderata trade-offs, and over-restriction (Figure \ref{fig:eval-adapt-iter3}). From the desiderata trade-offs scatterplot, we can see that a fitness function that upweights the interference metric accepts allocations with fewer exposed control units, at the cost of increased imbalance on covariates. The pairwise assignment correlation remains similar to complete randomization. 
\end{leftrule}

\begin{leftrule}
\textit{Final decision:} Ultimately, we proceed with the fitness function from Iteration 3 and the restriction rule used across iterations. We choose to prioritize mitigating interference over maximizing balance for two reasons. First, interference is typically very difficult, or even impossible, to measure and therefore hard to correct for in the analysis phase. Imbalance, on the other hand, can be addressed with standard covariate adjustment methods. Second, for the sexual assault prevention intervention in our simulated experiment, we particularly want to avoid false null findings because this can keep an intervention that helps protect girls' safety and well-being from being more widely implemented. Because interference often causes bias towards the null, we aim to prevent interference as much as possible. More generally, when deciding on a fitness function and restriction rule, study analysts may want to make methodological considerations, e.g. whether statistical corrections can be made during the analysis phase, and policy considerations, e.g. the consequences of a false negative or false positive finding.  
\end{leftrule}

\subsubsection{Pre-registration}
When the research team is satisfied with the choice of fitness function and restriction rule, they lock in the assignment mechanism $P_{IGR}$. They then pre-register the accepted allocations $\mathbf{z}^j$ in $\mathcal{Z}_{pool}$ that have $p_{\mathbf{z}_j} > 0$ and the associated probabilities $p_{\mathbf{z}}$. For the purpose of guaranteeing valid statistical inference, we only need to pre-register these accepted allocations and probabilities. However, it can also be helpful to pre-register the fitness function $f$ and the restriction rule $r$ as an informative reference for the design of future trials.
\begin{leftrule}
\paragraph{\textit{Pre-registration} for a Simulated Experiment}
We pre-register the set of 500 accepted allocations, each with equal probability $\frac{1}{500}$ of being drawn. Additionally, we pre-register the fitness function $f^{(\texttt{F+M})}$ and threshold restriction rule $r$.
\end{leftrule}

\subsubsection{Randomization}
Finally, the observed allocation $\mathbf{z}^{obs} \sim P_{IGR}$ is drawn and deployed.
\begin{leftrule}
\paragraph{\textit{Randomization} in a Simulated Experiment} We sample one of the 500 accepted allocations uniformly at random.
\end{leftrule}

\section{Analyzing IGR-Designed Experiments}\label{sec:igr-analysis}

In this section, we examine bias properties of a difference-in-means estimate applied to an IGR-designed experiment. We first discuss bias under the usual no-interference assumption, then expand the discussion to settings where interference is present. 

\subsection{IGR Bias Properties}\label{subsec:igr-bias}
\subsubsection{Unbiasedness Through the Mirror Property}
We show that IGR produces an unbiased difference-in-means estimate of the sample average treatment effect if there is no interference, if the IGR assignment mechanism satisfies the mirror property, and if in every accepted allocation, the number of treated and control individuals are equal. For a binary intervention, the mirror property is satisfied if $P(Z_i = 0) = P(Z_i = 1)$ for all $i = 1, \dots, N$. Suppose we enumerated \textit{all} possible allocations in the \textit{Enumeration} step. If we specify fitness function $f$ and restriction rule $r$ to be symmetric, such that $(r \circ f)(\mathbf{X}, \mathbf{z})) = (r \circ f)(\mathbf{X}, \mathbf{1} - \mathbf{z}))$, the mirror property will be satisfied by $P_{IGR}$. However, because we may not be able to enumerate all possible allocations, satisfaction of the mirror property is not guaranteed. Instead, we can update $P_{IGR}$ by ensuring that, for every allocation that is accepted, its mirror allocation is also accepted. Let $\tilde{\mathcal{Z}}_{pool} = \{1-\mathbf{z}  \mid \mathbf{z} \in \mathcal{Z}_{pool}\}$ denote the set of mirror allocations for each allocation in $\mathcal{Z}_{pool}$. The assignment mechanism $\tilde{P}_{IGR}$ satisfies the mirror property, where
\begin{align*}
    \tilde{P}_{IGR}(\mathbf{Z} = \mathbf{z}) = 
    \begin{cases}
    0 & \mathbf{z} \notin \mathcal{Z}_{pool} \cup \tilde{\mathcal{Z}}_{pool} \\
    0 & \mathbf{z} \in \mathcal{Z}_{pool} \cup \tilde{\mathcal{Z}}_{pool}, (r \circ f)(\mathbf{X}, \mathbf{z}) = 0 \\
    p_z & \mathbf{z} \in \mathcal{Z}_{pool} \cup \tilde{\mathcal{Z}}_{pool}, (r \circ f)(\mathbf{X}, \mathbf{z}) = 1
    \end{cases}
\end{align*}
\begin{proposition}\label{prop:bias}
    Assume that there is no interference and that the consistency property holds, so $Y_i = Y_i(1)Z_i + Y_i(0)(1-Z_i).$ If the assignment mechanism $\tilde{P}_{IGR}$ satisfies the mirror property and assigns an equal number of individuals to treatment and control in every allocation, then the difference-in-means estimate for the sample average treatment effect is unbiased.
\end{proposition}
\begin{proof}
    See Appendix \ref{app-subsec:proofs-unbiased-mirror}.
\end{proof}

\subsubsection{Bias in the Presence of Interference}
In an experiment with interference where the target estimand is the global average treatment effect, a design that satisfies the mirror property does not necessarily give an unbiased difference-in-means estimate because the interference itself introduces bias. To reduce interference bias, we may need to use a design that violates the mirror property. When the mirror property is violated, however, the individual's treatment assignment and potential outcome may no longer be independent, resulting in possible bias from confounding. It may nevertheless may be beneficial to break the mirror property so long as doing so produces a sufficient reduction in interference bias. 

In IGR, violation of the mirror property occurs when we choose an asymmetric fitness function and restriction rule pair to restrict the design. To understand why an asymmetric restriction can help to prevent interference, we consider a toy example. Suppose a subset of three or more individuals in the study are connected by a star-shaped network, such that there is a single central individual who is connected to every outer individual, and all outer individuals are only connected to this central individual. Suppose further that, for a unit assigned to the control-level, exposure to treatment occurs if at least two neighbors are treated. Then, in the allocation where the center individual is treated and the outer individuals are control, none of the control individuals are exposed. Conversely, in the mirror allocation where the center individual is control and the outer individuals are treated, the center control individual becomes exposed. The fitness function and restriction rule could be specified in such a way that, in order to rule out high-interference allocations, the latter allocation is rejected while the former is accepted. In Appendix \ref{app-subsec:proofs-bias-interference}, we derive conditions under which asymmetric restriction in IGR improves overall bias for a linear outcomes model with an additive direct effect and an additive neighborhood exposure spillover effect.

Note that, even if the fitness function and restriction rule are asymmetric, we can force the IGR assignment mechanism to satisfy the mirror property by incorporating all mirror allocations as ``accepted" allocations, regardless of whether they actually satisfy the criteria for acceptance. The analyst may consider enforcing the mirror property if the difference between the scores $f(\mathbf{X}, \mathbf{z})$ and $f(\mathbf{X}, \mathbf{1 - z})$ is small for the $\mathbf{z}$'s in the candidate pool of allocations. We empirically investigate the implications of including and excluding mirror allocations in Section \ref{sec:igr-effect-estimates}.

\subsection{Randomization Inference for IGR-Designed Experiments}
We take a design-based approach to inference, meaning we view the potential outcomes as fixed conditional on the study sample, so that the only randomness in the study arises through the randomness in the treatment assignment mechanism. In IGR specifically, randomness comes from the \textit{Randomization} step, where the observed allocation is sampled from the restricted assignment mechanism, $P_{IGR}$.  Randomization then forms the basis for inference, and we can perform a Fisher randomization test to evaluate null hypotheses. Specifically, we compute the value of the test statistic under the null hypothesis for different possible assignment vectors. With IGR, the empirical null distribution for the test statistic is derived over the set of accepted assignment vectors $\mathbf{z}^j$ in $\mathcal{Z}_{pool}$ that have $p_{\mathbf{z}_j} > 0$. We compare the observed value of the test statistic for $\mathbf{z}^{obs}$ to the empirically derived null distribution of the test statistic to obtain a \textit{p}-value.  

\section{How IGR Improves Effect Estimates}\label{sec:igr-effect-estimates}
Using IGR to design an experiment can lead to less biased, more precise effect estimates. We demonstrate these benefits with results from our simulated experiment (refer to Section \ref{subsec:sim-exp-interf} for a description of the simulation). In the \textit{Evaluation \& Adaptation} step in Section \ref{subsec:igr-step-by-step}, we chose the fitness function, $f^{(\texttt{F+M})}(\mathbf{z}, \mathbf{X}) = 0.75 \cdot m^{(\texttt{FracExpo})}(\mathbf{z}, \mathbf{X}) + 0.25 \cdot m^{(\texttt{Mahalanobis})}(\mathbf{z}, \mathbf{X})$, to apply to the design of the simulated experiment. Because the interference metric $m^{(\texttt{FracExpo})}$ is asymmetric, the resulting restricted assignment mechanism does not generally satisfy the mirror property unless we incorporate all mirror allocations, including those that would not be accepted by the restriction rule (refer to Section \ref{subsec:igr-bias} for a discussion on the mirror property and estimation bias). To understand the implications of satisfying versus violating the mirror property, we examine effect estimates both with and without the inclusion of mirror allocations. We additionally assess how the estimation properties change when we apply different weights to the interference and balance metrics in the fitness function; below, we refer to the weight applied to the interference metric, $m^{(\texttt{FracExpo})}$, as $w_{Interference}$. We also show how, when the interference metric is upweighted in the design phase, the incurred covariate imbalance can be corrected for in the analysis phase through covariate adjustment. Results are reported based on five repeatedly drawn samples of the covariate and outcome data.

\begin{figure}[ht!]
    \centering
    \includegraphics[width=0.65\linewidth]{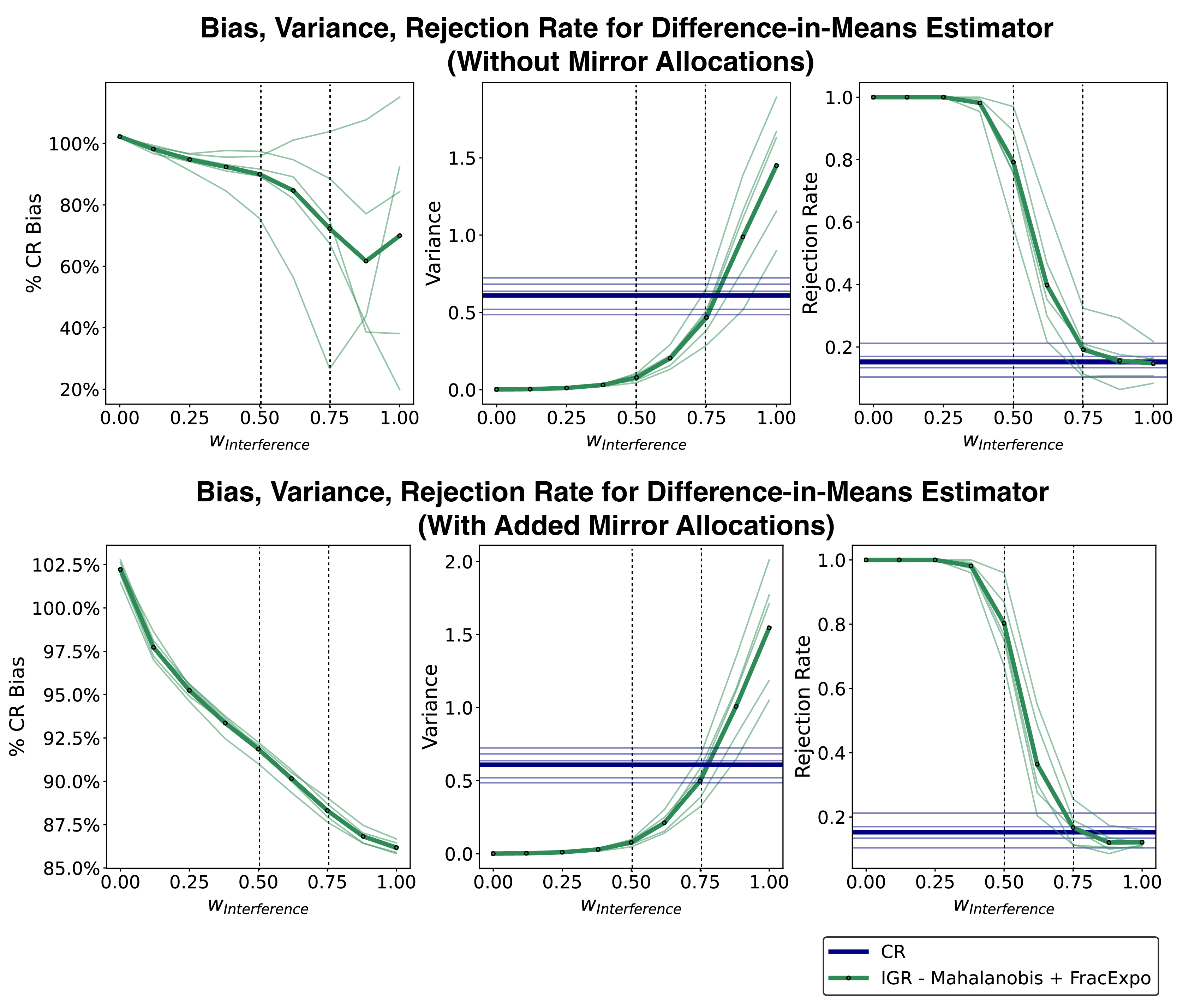}
    \caption{Bias, variance, and rejection rate of the difference-in-means estimator in an IGR design versus an unrestricted complete randomization (CR) design when mirror allocations are excluded (top) and when mirror allocations are included (bottom). Bias is plotted as a percentage relative to the bias of a CR design.}
    \label{fig:bias-var-rr-dim}
\end{figure}

When mirrors are included and $w_{Interference} = 0.75$, IGR reduces the absolute bias of the difference-in-means estimate ($88.31 \pm 0.56 \%$ of the absolute bias of an unrestricted design)(Figure \ref{fig:bias-var-rr-dim}). As the magnitude of the weight applied to the interference metric increases, we observe more bias reduction ($86.18 \pm 0.37\%$ of the absolute bias of an unrestricted design when $w_{Interference} = 1$). At small values of $w_{Interference}$, IGR decreases the variance of the difference-in-means estimate (e.g., variance at $w_{Interference} = 0.125$ is $0.003 \pm 0.0007$ compared to $0.61 \pm 0.10$ for an unrestricted design). Increasing $w_{Interference}$ leads to variance inflation (e.g. variance at $w_{Interference}=0.875$ is $1.01 \pm 0.28$). When $w_{Interference} = 0.75$, leaving mirrors out leads to a large reduction in absolute bias for some data samples (e.g., $26.67 \%$ of the absolute bias of the unrestricted design) and no reduction for others (e.g., $103.89\%$ of the absolute bias of the unrestricted design). This suggests that, in our example, including mirror allocations helps to make the estimation bias less dependent on the particular sample of individuals in the study. 

\begin{figure}[ht!]
    \centering
    \includegraphics[width=0.65\linewidth]{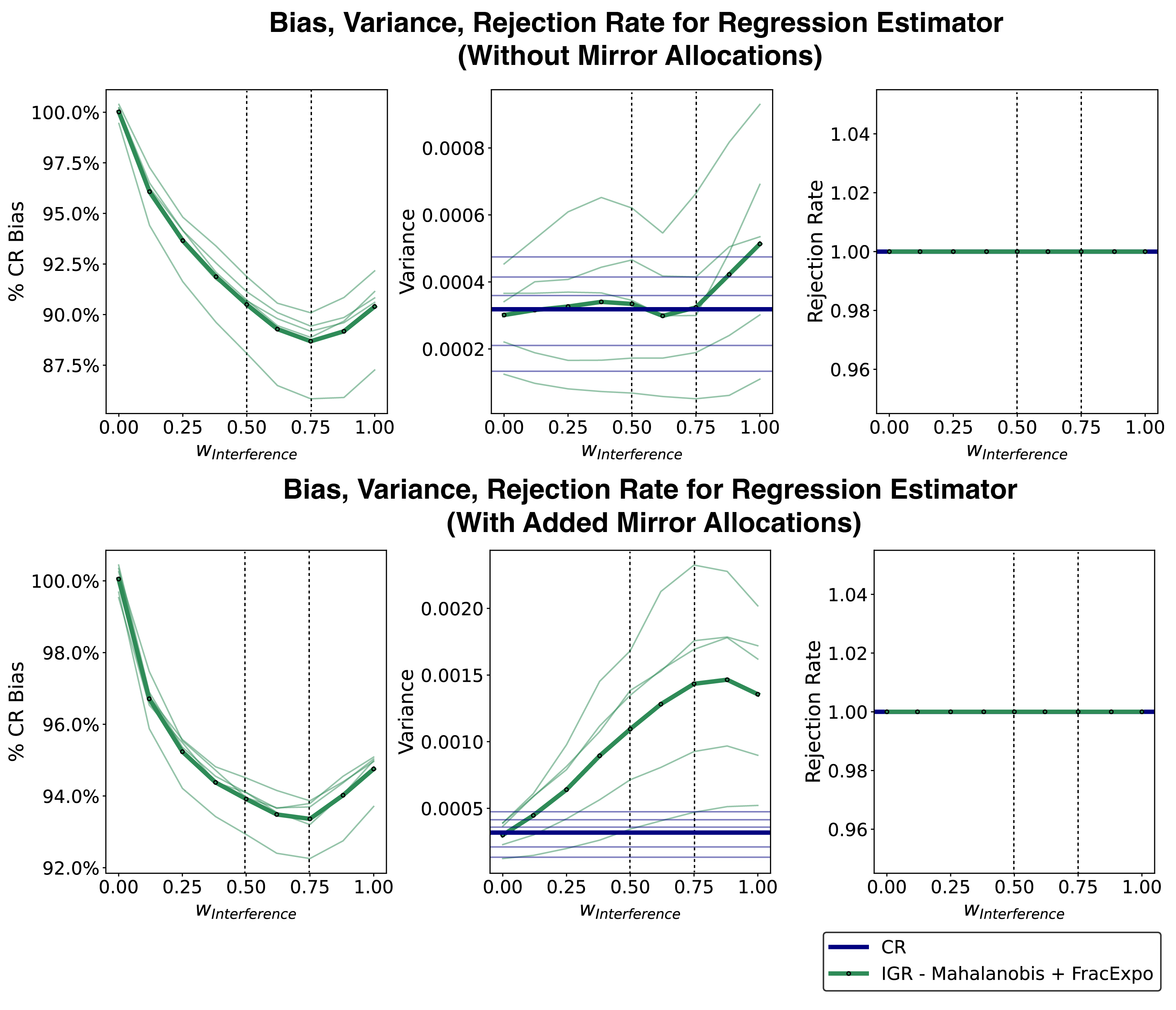}
    \caption{Bias, variance, and rejection rate of the linear regression estimator in an IGR design versus an unrestricted complete randomization (CR) design when mirror allocations are excluded (top) and when mirror allocations are included (bottom). Bias is plotted as a percentage relative to the bias of a CR design.}
    \label{fig:bias-var-rr-linreg}
\end{figure}

Instead of a difference-in-means estimator, we can also fit a linear regression model that adjusts for observed baseline covariates. As mentioned in the \textit{Evaluation \& Adaptation} step in Section \ref{subsec:igr-step-by-step}, using this estimator can be helpful when we choose to upweight the interference metric and downweight the balance metric in the design phase. As expected, the regression estimator substantially reduces variance compared to the difference-in-means estimator, particularly when $w_{Interference}$ is large (Figure \ref{fig:bias-var-rr-linreg}). When mirror allocations are excluded, the bias in the regression estimate is more stable across data samples compared to the difference-in-means estimate. On the other hand, when mirror allocations are included, the regression estimate generally provides less bias reduction than the difference-in-means estimate ($93.36 \pm 0.67\%$ of the absolute bias of an unrestricted design at $w_{Interference}=0.75$).

From the above, we reason that, when using a a difference-in-means estimator, it is beneficial to include mirror allocations and to not over-weight the interference metric, e.g. above 0.5. If using the regression estimator, however, mirror allocations can be excluded, and a larger $w_{Interference}$ can be applied without needing to worry about the variance of the estimator. We do note, however, that a difference-in-means estimator is generally less amenable to \textit{p}-hacking, which is one factor to consider when choosing between these two estimators. 

\vspace{-10pt}

\section{Discussion}\label{sec:disc}
We introduced Inspection-Guided Randomization as a flexible framework for performing restricted randomization with diverse design desiderata and used illustrative examples to showcase its usefulness in different experimental settings. IGR is distinct from prior restricted randomization frameworks along several fronts. It allows for multiple design desiderata, including and beyond balance, thus expanding the use cases for restricted randomization; we showed this with an example of an experiment with interference in Section \ref{sec:igr-detail} and a group formation experiment in Appendix \ref{app-sec:group-form}. IGR also separates the process of enumerating candidate allocations -- which can be done with an efficient, directed search -- from the act of randomization; we explicitly make enumeration and randomization separate steps in the IGR framework. In addition, IGR has a built-in iterative loop for evaluating and adapting the restricted assignment mechanism; we show this in action in Section \ref{subsec:igr-step-by-step}, using the help of visual tools described in Appendix \ref{app-sec:eval-adapt-elab}. Lastly, IGR enforces pre-registration of the accepted allocations, thus promoting transparency and reproducibility; in Appendix \ref{app-sec:phack-rerand}, we show that, without pre-registration, it is easy to inflate the Type I error rate above the nominal level.

The core motivation for IGR, as for any restricted randomization, is to prevent the possibility of choosing a bad randomization allocation that leads to an effect estimate that deviates dramatically from the true value of the estimand. For obviously bad allocations, there is little reason not to discard them; this has been well-understood since the early days of the experimental design literature. In a conversation between Cochran and Fisher (recounted by Rubin and Cochran), Fisher is reported to have said that, confronted with a randomized allocation that ``exhibited substantial imbalance on a prognostically important baseline covariate," he would ``of course" rerandomize if the experiment had not started yet \citep{Rubin2008-va}. 

Historically, the focus of restricted randomization has been on achieving covariate balance in two-arm experiments that compare a single treatment intervention to a control or status quo. However, some interventions, often ones that are behavioral, need to be evaluated through more complicated restricted designs. Researchers across several disciplines, including medicine, education, psychology, and computer science, have recognized this need and have independently applied principles of the IGR framework to designing experiments, albeit without the formal structure of IGR \citep{Murnane2023-vy, Cho2021-pf, Cho2022-oy, Kizilcec2022-ul, Zahrt2023-jd}. \cite{Cho2021-pf} and \cite{Cho2022-oy}, for example, used a form of restricted randomization to achieve balance on prognostic covariates in trials with more than two arms. These existing real-life applications suggest that IGR is conceptually straightforward and intuitive, and they also point to how an explicit IGR framework can help to bring consistency to existing efforts to expand the scope of restricted randomization. 

In this work, we showed how the IGR framework can be used to design experiments in two settings where existing restricted randomization frameworks are not immediately applicable. In our first example, we applied IGR to the design of experiments with interference. Experimental designs that seek to mitigate interference often do so through a form of cluster randomization, either over naturally existent clusters or over algorithmically detected ones \citep{Ugander2013-tf, Ugander2023-zk}. Without IGR, the study analyst chooses between randomizing at the more aggregated level to limit interference, thus sacrificing power and covariate balance, or randomizing at the more granular level to increase efficiency, thus permitting more interference bias. With IGR, the analyst can construct fitness functions that combine both interference and balance metrics, thus reducing interference bias while maintaining efficiency. IGR can also be used even if we have limited knowledge of interference structure and need to rely on an approximate measure of interference. While we explored a specific setting where interference occurred via neighborhood exposure in a network of interactions, IGR can easily be applied to other interference structures by modifying the inspection metric used to measure the extent of anticipated interference.

In Appendix \ref{app-sec:group-form}, we further show how IGR can be applied to the design of group formation experiments, where effect sizes may exhibit dependence on group composition. Recent work has highlighted that peer context should be taken into consideration as a driver of treatment effect heterogeneity for behavioral and psychological interventions \citep{Yeager2019-kr, Walton2020-tr,kizilcec2020scaling, kizilcec2017eight}. Understanding how an intervention depends on peer context has important implications for understanding causal mechanisms, for devising policies that target certain contexts, and for designing concurrent interventions that can be used to manipulate contexts. 

We call out the \textit{Evaluation \& Adaptation} step of IGR as an especially critical component of the framework. In some sense, the inclusion of the \textit{Evaluation \& Adaptation} step is a recognition that the study analyst is not all-knowing and hence may not select a good fitness function and restriction rule in their first attempt. In this work, we highlighted three issues to consider in \textit{Evaluation \& Adaptation}: desiderata trade-offs, discriminatory power, and over-restriction. We introduced visual tools to help with assessing each of these issues (described in detail in Appendix \ref{app-sec:eval-adapt-elab} and applied in Section \ref{subsec:igr-step-by-step}). For example, in our simulated experiment, the analyst may initially choose to assign equal weight to the interference metric and the balance metric (Figure \ref{fig:eval-adapt-iter2}). After considering the risk of a false null, however, the analyst may choose to modify the weighting scheme to instead upweight the interference metric. These iterative refinements push the design towards one that is better and more likely to yield credible inference. As mentioned in Related Works (Section \ref{sec:rel}), an emphasis on detecting and correcting flawed restricted assignment mechanisms is missing from standard restricted randomization frameworks such as re-randomization.

Fundamentally, IGR compels the study analyst to invest time and care in the design phase of the experiment. The \textit{Specification} and \textit{Evaluation \& Adaptation} steps in particular require considerable thoughtfulness, from specifying the study's design desiderata, to developing the inspection metrics used to measure desiderata satisfaction, to constructing and evaluating the fitness function and restriction rule that characterize the restricted assignment mechanism. Attentiveness to design can produce large payoffs in the bias and precision of effect estimates. While using a tailored estimator in the analysis phase can sometimes produce similar payoffs, focusing efforts on the design phase has additional merits. For one, even the most clever estimators may not be able to salvage a study that was performed under an extremely poor choice of randomization allocation. In addition, as previously mentioned, the \textit{Pre-registration} step precludes statistical hacking, as long as the analyst is held accountable to doing randomization inference over a pre-registered space of randomization allocations (see Appendix \ref{app-sec:phack-rerand} for an illustration of Type I error rate inflation when the analyst is not held accountable to a pre-registered design and analysis plan). 

There are several interesting directions for future work on IGR. Since the fitness function in IGR is hand-crafted, it may be useful to develop theory on how similar the specified fitness function needs to be to the true conditional mean-squared error in order for IGR to produce better effect estimates than naive randomization strategies. When the fitness function is not a faithful proxy, then analysts may choose to use a non-IGR design. Alternatively, analysts could choose to improve the fitness function by using additional sources of data and domain knowledge. If experiments consist of multiple waves or are preceded by a pilot study, then there is opportunity to use information collected from the prior wave or from the pilot study to refine the fitness function used in the next wave or in the actual trial. In an experiment with anticipated interference, for example, researchers could use a pilot study to estimate the interference network structure within a subsample of the study participants, then extrapolate this network to the entirety of the study sample. Fitness functions for new experiments could potentially also leverage data from prior historical experiments that evaluated different but comparable interventions in similar settings. Lastly, to make IGR as accessible as possible by study analysts, good software for implementation will be useful; this will likely include a dashboard interface that allows users to easily perform each of the steps of IGR, including interacting with visualizations.  


\clearpage
\appendix
\numberwithin{equation}{section}
\counterwithin{table}{section}
\counterwithin{figure}{section}

\section{Applications of Tightening, Re-randomization, and Constrained Randomization Use Balance Criteria}

    \begin{table}[ht!]
    \footnotesize
    \centering
    \begin{tabular}{p{6cm}p{3cm}p{6cm}}
    \toprule
    \textbf{Paper} & \textbf{Randomization strategy} & \textbf{Criteria} \\
    \midrule
    \cite{Tukey1993-no} & Constrained & Balance in treatment group sizes within covariate strata \\
    \cite{Raab2001-to} & Constrained & L2 balance \\
    \cite{Moulton2004-sj} & Constrained & Absolute difference in mean covariate value, geographic balance \\
    \cite{Morgan2012-mb} & Re-randomization & Mahalanobis distance \\
    \cite{De_Hoop2012-bm} & Constrained & L2 balance \\
    \cite{Li2016-zf} & Constrained & L2 balance, L1 balance \\
    \cite{Branson2016-fm} & Re-randomization & Mahalanobis distance \\
    \cite{Li2017-yg} & Constrained & L2 balance, L1 balance \\
    \cite{Basse2018-ii} & Re-randomization & Network balance (e.g. average neighborhood size) \\
    \cite{Li2018-gh} & Re-randomization & Mahalanobis distance \\
    \cite{Ciolino2019-ms} & Constrained & Kruskal-Wallis test p-value \\
    \cite{Krieger2020-hl} & Re-randomization & Absolute difference in mean covariate \\
    \cite{Watson2021-ob} & Constrained & Trace of the between-arm covariance matrix \\
    \cite{Zhou2021-rq} & Constrained & Maximum pairwise L2 balance, maximum pairwise Mahalanobis distance \\
    \cite{Nordin2022-jw} & Re-randomization & Mahalanobis distance \\
    \cite{Kapelner2021-fy} & Re-randomization & Mahalanobis distance \\
    \bottomrule
    \end{tabular}
    \caption{\textbf{Examples of re-randomization and constrained randomization in the literature.} Re-randomization refers to the strategy where allocations are sampled one at a time until the first allocation that satisfies the criterion is drawn. Constrained randomization refers to the strategy where a set of candidate allocations are sampled simultaneously, then filtered down to a set of accepted allocations from which the final allocation is drawn. All of these works use a type of balance criterion (including Basse and Airoldi, 2018 who use a balance criterion based on network properties in order to account for network-correlated outcomes).}
    \label{tab:rerand-lit-review}
\end{table}

\clearpage
\section{General Simulation Pipeline}\label{app-sec:sim-pipe}
\begin{algorithm} [!ht]
\caption*{\textbf{Simulation Pipeline}} \label{alg:sim-pipeline}
\begin{enumerate}
    \item Generate baseline covariates.
    \item ``Design" the experiment, either using IGR or using a benchmark design strategy. 
    \begin{enumerate}
        \item \textit{For IGR:} Enumerate a pool of $M$ candidate allocations and apply a thresholding restriction rule where the threshold is defined as the $m^{th}$ smallest score. Specifically, keep the top $\frac{m}{k}$ scoring allocations and possibly their corresponding $\frac{(k-1)\cdot m}{k}$ mirror allocations (regardless of score) as acceptable allocations for the experiment. 
        \item \textit{For benchmark design strategies:} Generate $m$ possible allocations under the benchmark design. While this may not be inherently necessary for executing the design itself, doing so facilitates comparability of the design with IGR. 
    \end{enumerate}
    \item ``Run" the experiment under each of the $m$ accepted allocations.
    \begin{enumerate}
        \item For each allocation, generate the observed outcomes. Then, estimate the target treatment effect estimand and use a randomization test to compute an exact \textit{p}-value for a test of no effect.
        \item Estimate bias, mean-squared error, and rejection rate using the treatment effect estimates and \textit{p}-values across the $m$ re-runs.
    \end{enumerate}
\end{enumerate}
\end{algorithm}

\clearpage
\section{Data Generation for Experiment With Interference}\label{app-sec:interference-dgp}
We use GoogleEarth to determine an approximately central coordinate for each of the five settlements, then sample school coordinates in a grid around the settlement center. We generate covariates $\tilde{\mathbf{x}}_i = \begin{pmatrix}
    \tilde{x}_{i1}, \tilde{x}_{i2}
\end{pmatrix}^T $for each individual as follows:
\begin{align}
    \begin{split}
    \tilde{x}_{ij} &\sim N(\mu_{s(i), j}, \sigma_{individual, j}) \\
    \mu_{s(i), j} &\sim N(\mu_{t((s(i)), j}, \sigma_{school, j}) \\
    \text{for } j &= 1, 2
    \end{split}
\end{align}
where $s(i)$ denotes the school that $i$ attends and $t(s(i)) \in \{1, 2, 3, 4, 5\}$ denotes the settlement in which $s(i)$ is located. We manually set $\mu_{t, j}$ to fixed values for each settlement $t \in \{1, 2, 3, 4, 5\}$ and each $j \in \{1, 2\}$. Similarly, we also manually set $\sigma_{individual, j}, \sigma_{school, j}$ for each $j \in \{1, 2\}$. To introduce homophily, such that neighbors in the interference network are more similar in their covariates, we define $\mathbf{x}_i$ to be the sum of $\tilde{x}_i$ and $\tilde{x}_j$ for all $j$ that are $i$'s neighbors in the network. Formally,
\begin{align}
    \mathbf{x}_i = \mathbf{A} \tilde{\mathbf{X}}
\end{align}
where $\mathbf{A}$ is an adjacency matrix for the interference network and $\mathbf{\tilde{X}} = \begin{pmatrix}\tilde{x}_1 & \tilde{x}_2 & \cdots & \tilde{x}_n \end{pmatrix}^T$. We sample edges for the underlying interference network using the following model:
\begin{align}
    P(A_{ij} = 1) &= \begin{cases}
        \theta^{w} & s(i) = s(j) \\
        \theta^{b1} \cdot \tilde{d}(i,j)^{-\gamma} & s(i) \neq s(j), t(i) = t(j) \\
        \theta^{b2} \cdot \tilde{d}(i,j)^{-\gamma} & s(i) \neq s(j), t(i) \neq t(j) \\
    \end{cases} \\
    \tilde{d}(i, j) &= \frac{||\mathbf{l}_{s(i)} - \mathbf{l}_{s(j)}||_2^2}{\min_{(i, j) \in [N] \times [N], s(i) \neq s(j)} ||\mathbf{l}_{s(i)} - \mathbf{l}_{s(j)}||_2^2}
\end{align}
where $A_{ij}$ is the $(i,j)$-th entry of the adjacency matrix, $s(\cdot)$ denotes school membership, $t(s(\cdot))$ denotes settlement membership, and $\mathbf{l}_{s(\cdot)}$ denotes the school's 2-dimensional latitude and longitude coordinates. $\theta^{w}$, $\theta^{b1}$, and $\theta^{b2}$ are scalar factors. We distinguish between $\theta^{b1}$ and $\theta^{b2}$ because individuals who attend different schools within the same settlement may be more likely to be connected than individuals who attend different schools in different settlements, even if the schools are the same distance apart.

To generate the observed outcome $Y_i^{obs}$ for individual $i$ under global treatment assignment vector $\mathbf{z}^{obs} \in \{0, 1\}^N$, we use the following model, with $\tau$ capturing the direct effect of an individual's own treatment and $\delta$ capturing the additive spillover effect:
\begin{align}
    Y_i^{obs} = \mathbf{X}\boldsymbol{\beta} + \tau z_i + \delta \mathbf{1} \left\{ \sum_{j \in \mathcal{N}(i)} z_j > q \cdot |\mathcal{N}(i)| \right\}(1-z_i)
\end{align}
Note that, under this model, the global average treatment effect $\tau_{TATE}$ is exactly equal to $\tau$. Furthermore, if we use an experimental design that assigns a fixed number of units $N_c$ to control, then the bias of a difference-in-means estimator is 
\begin{align*}
    E[\hat{\tau} - \tau] = -\delta \cdot \frac{1}{N_c} \sum_{i=1}^{N} \mathbb{E} \left[ \mathbf{1} \left\{ \sum_{j \in \mathcal{N}(i)} z_j > q \cdot |\mathcal{N}(i)| \right\} (1-z_i) \right]
\end{align*}
where the expectation is taken over all possible allocations. Thus, in order to reduce interference bias, we should rule out allocations with many exposed control units.

We run simulations for
($\boldsymbol{\mu}_{\cdot 1}, \boldsymbol{\mu}_{\cdot 2}) = 
        \left(\begin{pmatrix}
            0.25 & 0 & 0.05 & 0 & 1
        \end{pmatrix},
        \begin{pmatrix}
            0.25 & 0.75 & 0 & 0.25 & 0
        \end{pmatrix}\right)$; \\
$(\sigma_{individual, j}, \sigma_{school, j}) = (0.1 \cdot sd(\mu_{\cdot j}), 0.2 \cdot sd(\mu_{\cdot j}))$;
$\gamma = 0.50$; 
($\theta^{w}, \theta^{b1}, \theta^{b2}) = (0.2, 0.1, 0.01)$;
$\boldsymbol{\beta} = \begin{pmatrix}
    1 & 1
\end{pmatrix}^T$; $q=0.25$; and
$\tau = \delta = 0.3 \cdot SD(\mathbf{X}\boldsymbol{\beta})$. 

\clearpage
\section{Toy Example Visualizations for the \textit{Evaluation \& Adaptation} Step of IGR}\label{app-sec:eval-adapt-elab}

When applying the \textit{Evaluation \& Adaptation} step of IGR to our simulated example (see Section \ref{subsec:igr-step-by-step}), we used visualizations to help check for discriminatory power, desiderata trade-offs, and over-restriction. In this section, we provide a more detailed description and rationale for each type of visualization used. We use synthetic toy examples to assist in our explanations.

\subsection{Discriminatory Power}
\begin{figure}[ht!]
    \centering
    \includegraphics[width=0.5\linewidth]{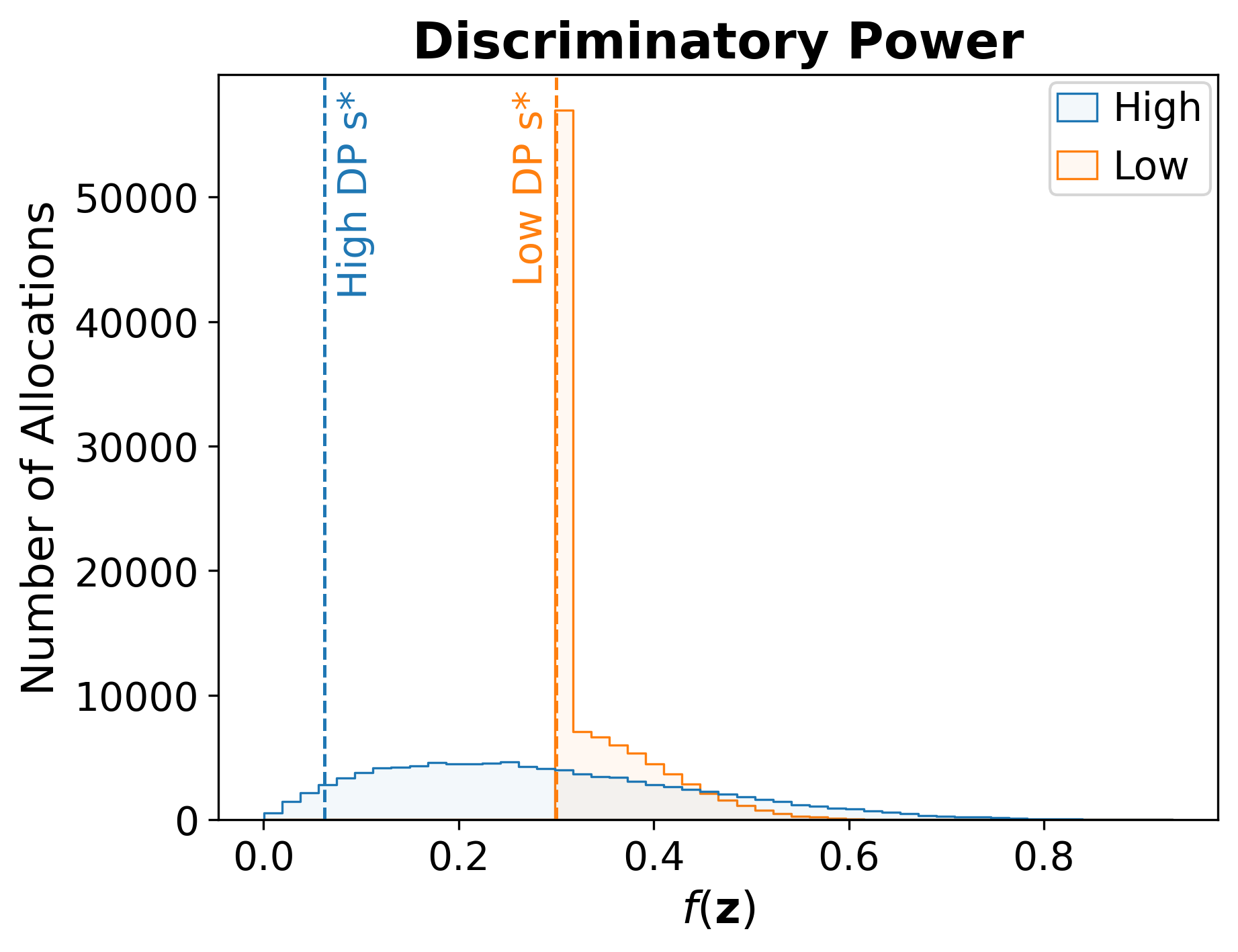}
    \caption{Histograms of fitness function scores across candidate allocations for a fitness function with low discriminatory power (orange) and high discriminatory power (blue). Note that smaller values of $f(\mathbf{z})$ are considered more desirable.}
    \label{app-fig:dp}
\end{figure}

We highlighted discriminatory power as one check to make during the \textit{Evaluation \& Adaptation} step of IGR. To perform this check, we can plot a histogram of the fitness scores across all candidate allocations. In Figure \ref{app-fig:dp}, we show two example distributions of fitness scores that one might encounter. A distribution with poor or low discriminatory power is one where many of the allocations have the same or similar score, which is easily identifiable in a histogram. A distribution with high discriminatory power is one where the scores are more uniformly distributed, such that the better-scoring allocations are clearly distinguished from the worse-scoring allocations.

\subsection{Desiderata Trade-offs}
We also highlighted desiderata trade-offs as an important check to make in the \textit{Evaluation \& Adaptation} step. When there are two desiderata -- and two corresponding inspection metrics for measuring them -- we can use two-dimensional histograms and scatterplots to help with performing this check.

\begin{figure}[ht!]
    \centering
    \includegraphics[width=\linewidth]{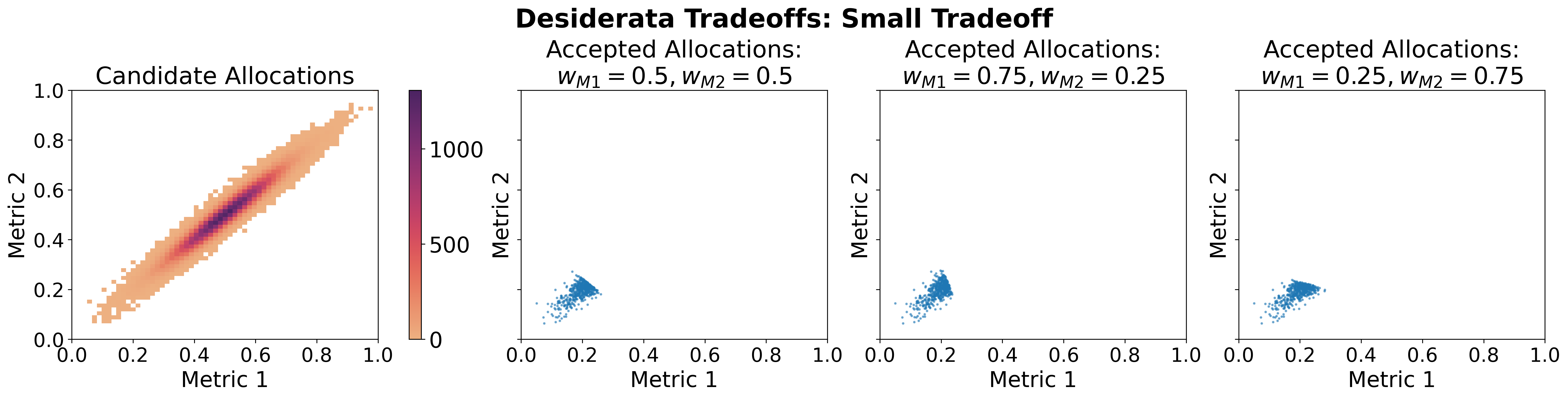}
    \caption{(left) Histogram of inspection metric values across candidate allocations. Notice that the two metrics are positively correlated, meaning smaller and better values of one metric correspond to smaller and better values of the other. (right) Scatterplots of inspection metric values across accepted allocations under fitness functions that apply different weights to the metrics. When the trade-off between the metrics is small, the choice of weighting scheme does not have a large effect on the set of accepted allocations.}
    \label{app-fig:dt-small}
\end{figure}

\begin{figure}[ht!]
    \centering
    \includegraphics[width=\linewidth]{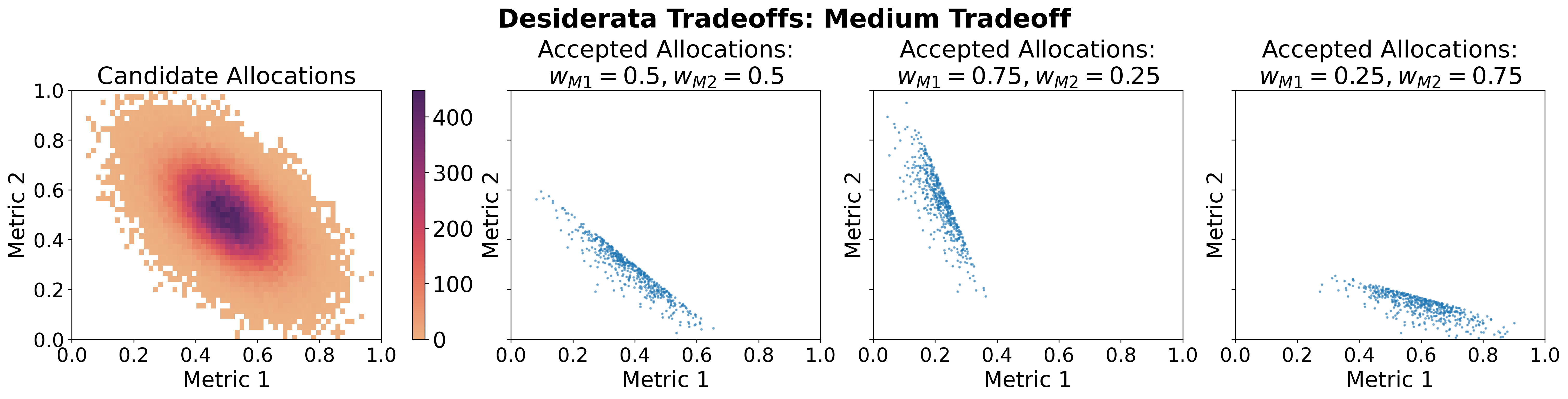}
    \caption{(left) Histogram of inspection metric values across candidate allocations. Notice that the two metrics are somewhat negatively correlated, meaning smaller and better values of one metric tend to correspond to larger and worse values of the other. (right) Scatterplots of inspection metric values across accepted allocations under fitness functions that apply different weights to the metrics. By applying a greater weight to Metric 1 ($w_{M1} = 0.75, w_{M2} = 0.25$), we prioritize accepting allocations with a smaller value of Metric 1, at the cost of permitting some allocations with a larger value of Metric 2. An analogous trade-off occurs when we apply greater weight to Metric 2 ($w_{M1} = 0.25, w_{M2} = 0.75$).}
    \label{app-fig:dt-med}
\end{figure}

\begin{figure}[ht!]
    \centering
    \includegraphics[width=\linewidth]{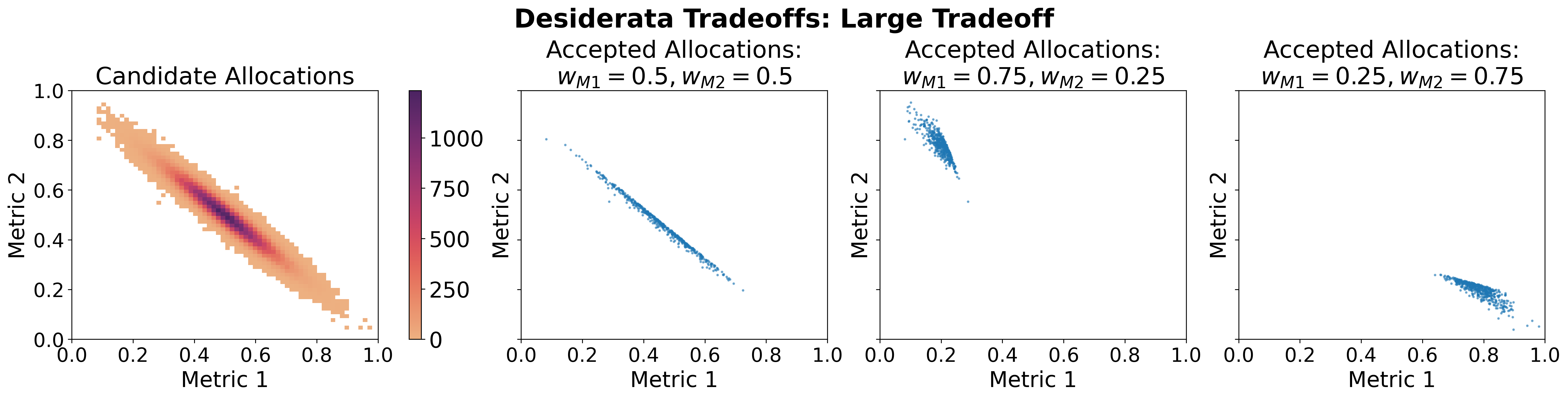}
    \caption{(left) Histogram of inspection metric values across candidate allocations. Notice that there is a strong negative correlation between the two metrics, meaning the best values of one metric tend to correspond to the worst values of the other. (right) Scatterplots of inspection metric values across accepted allocations under fitness functions that apply different weights to the metrics. When the trade-off is large, we may not be able to avoid accepting some allocations with a small value of one metric and a large value of the other, no matter what kind of weighting scheme we use.}
    \label{app-fig:dt-large}
\end{figure}

In the two-dimensional histogram, we plot the distribution of pairs of inspection metric values across all candidate allocations. This helps to reveal the extent to which desiderata trade-offs exist in the candidate space. In Figures \ref{app-fig:dt-small} - \ref{app-fig:dt-large}, we show three example histograms that suggest small, medium, and large trade-offs, respectively. When the trade-off between the metrics is non-existent or small, then smaller values of one metric also tend to correspond to smaller values of the other metric. On the other hand, for medium or large trade-offs, smaller values of one metric tend to correlate with larger values of the other metric.

In the scatterplot, we plot pairs of inspection metric values across the \textit{accepted} allocations. This helps to reveal the extent to which the chosen fitness function and restriction rule ``navigate" the trade-offs in the candidate allocation space. In Figures \ref{app-fig:dt-small} - \ref{app-fig:dt-large}, we show accepted allocations under a fitness function that linearly combines the two inspection metrics. When there are medium to large trade-offs between desiderata, then the set of accepted allocations is more sensitive to the weights that are used in the linear combination.

\subsection{Over-restriction}
\begin{figure}[ht!]
    \centering
    \includegraphics[width=0.5\linewidth]{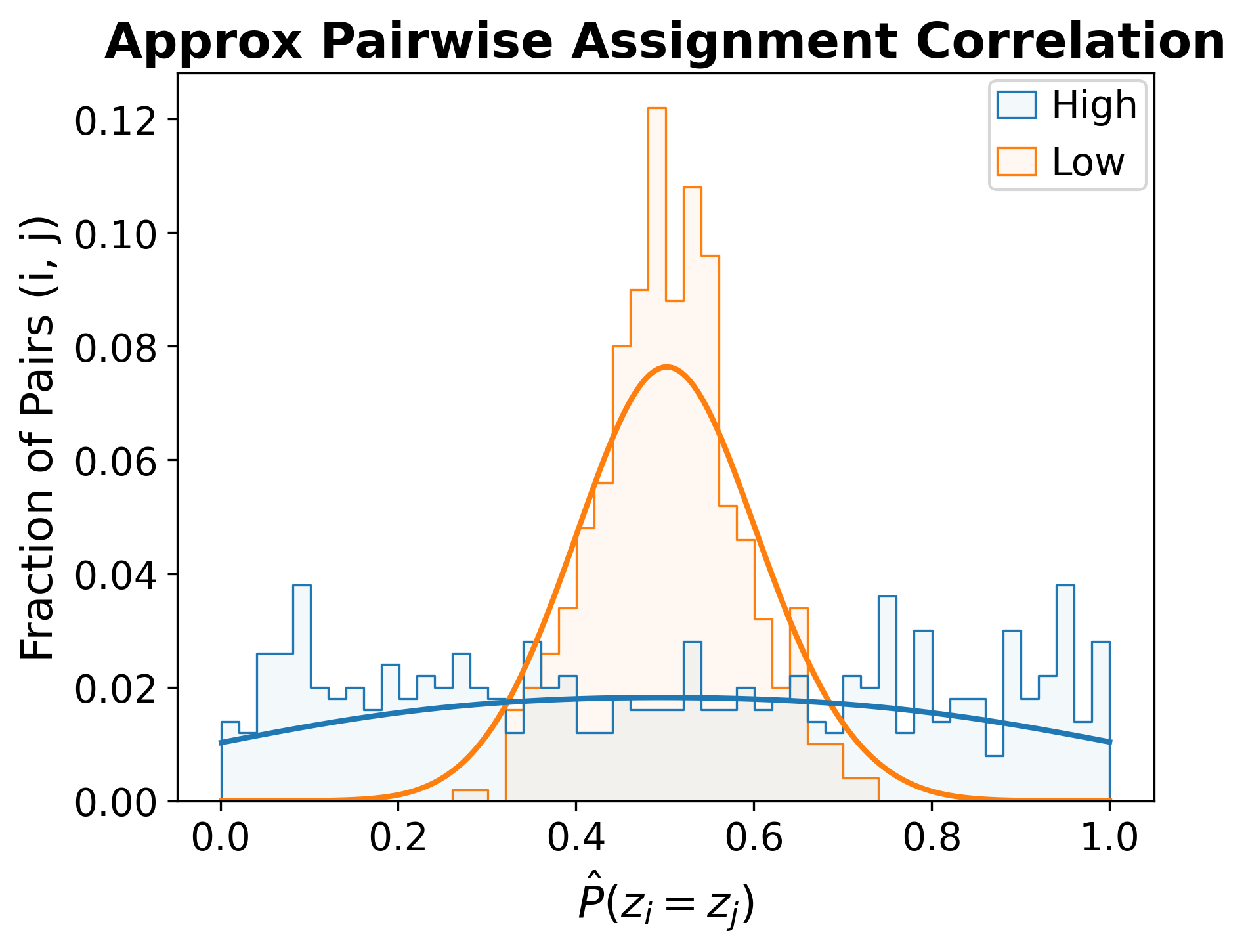}
    \caption{Histograms of approximate pairwise assignment correlation to check for over-restriction. For high pairwise assignment correlation (blue), the histogram has mass closer to 0 and to 1, indicating that there are many pairs of units $i, j$ that are nearly always assigned to different or to the same arm. For low pairwise assignment correlation (orange), pairs of units are assigned to the same arm around half the time.}
    \label{app-fig:or}
\end{figure}

Lastly, we highlight checking for over-restriction. In particular, we examine the extent to which there is pairwise assignment correlation across the accepted allocations. As mentioned in the \textit{Evaluation \& Adaptation} step in Section \ref{subsec:igr-step-by-step}, pairwise assignment correlation can reduce power. As an estimate of pairwise correlation, for each pair of units $i, j$, we calculate the fraction of accepted allocations where $i, j$ are assigned to the same arm, $\hat{P}(z_i = z_j) = \sum_{\mathbf{z} \in \mathcal{Z}_{pool}, p_{\mathbf{z}} > 0} \mathbf{1}\{z_i = z_j \}$. We then plot a histogram of these fractions. When there is high assignment correlation, then we expect the histogram to have mass at values closer to 0 and 1, which correspond to a pair $i, j$ never being assigned to the same arm and always being assigned to the same arm, respectively. When there is little to no assignment correlation, the histogram should be largely concentrated around 0.5. In Figure \ref{app-fig:or}, we show an example of low and high pairwise assignment correlation.

\clearpage
\section{IGR for a Simulated Group Formation Experiment}\label{app-sec:group-form}

In this section, we present another example use case for IGR where prior restricted randomization strategies have not been applied.

\subsection{Simulated Group Formation Experiment}
\paragraph{Group Formation Experiments for Composition-Dependent Effects}
In a group formation experiment, the challenge is to minimize differences in most baseline covariates, while creating large contrasts in a salient covariate of interest. Individuals are randomized into different groups, typically to construct groups of certain compositions that are believed to have different responses to the intervention \citep{Basse2024-sw}. For example, a group may be the set of college freshmen assigned to the same dorm room, and composition may be the number of individuals in a room who were ``high academic achievers" in high school \citep{Sacerdote2001-bg}. We consider group formation trials where composition is defined in terms of a categorical salient attribute (e.g., high or low prior academic achievement) and is count-based (e.g., number of individuals with high academic achievement). 

We illustrate how to use IGR in the design of a group formation experiment where we are interested in evaluating how group composition influences the effect of a group-level intervention. Various kinds of inter-group comparisons may be of interest, including between treated and control groups with the same compositions (to target the treatment effect under a fixed composition) and between treated and treated groups with different compositions (to target the effect of the composition under a fixed treatment). In addition, we may also want to make difference-in-differences comparisons between the treatment effect under one composition and the treatment effect under a different composition.

The design goals in this setting are two-fold. First, we would like to sort individuals into groups such that certain pre-specified group compositions are realized. We focus in particular on the case where composition is characterized by the proportion of the group that possesses a particular salient attribute. Second, we would like to maintain balance on all covariates other than those used to define composition. A key point of this vignette is to highlight that in the \textit{Enumeration} step of IGR, we can use different methods to sample allocations. Here, we sample from a blocking-based group formation assignment mechanism.

\paragraph{Simulation Set-up}
We simulate the study data set to encode theory-based covariance between student characteristics (age, gender, and college major), latent traits (ability and confidence), and course performance (homework score and final exam grade). In particular, course performance is assumed to vary with students’ latent ability and confidence, which in turn depend on age, gender, and major. Ability is modeled to increase with age and is higher for some majors than other majors. In the context of STEM majors where the majority of students are men, we model confidence to be higher among men than women. We also model higher confidence for certain STEM-related majors.

Let $\mathbf{x}_i = \begin{pmatrix} age_i & gender_i & major_i & hw_i & exam_i \end{pmatrix}^T$ denote observed baseline covariates for individual $i$ and $\mathbf{u}_i = \begin{pmatrix} ability_i & confidence_i \end{pmatrix}^T $ denote latent, unobserved covariates. To generate $gender_i$, we randomly set half of the study sample to be men. Although having a fixed gender composition in the study population is not strictly necessary, it makes it easier to implement a group formation strategy that assembles groups with certain gender compositions. In a real-world setting, we could construct a study population with a fixed gender ratio by using an enrollment approach that intentionally targets this ratio. We use the following data-generating process for the remaining covariates: 
\begin{small}
    \begin{align*}
    age_i &\sim \text{Unif}(19, 25) \\
    major_i &\sim \{1, 2, 3\} \text{ with probability } \{0.5, 0.3, 0.2\} \\
    ability_i &= (age_i - \bar{age}) / \sigma_{age} - 0.5 \cdot \mathbf{1} \{gender_i = 1 \} + \mathbf{1} \{major_i = 1 \} + \varepsilon \\
    confidence_i &= \mathbf{1} \{gender_i = 1 \} + \mathbf{1} \{major_i = 2\} + \varepsilon \\
    hw_i &= ability_i + \epsilon \\
    exam_i &= ability_i + confidence_i + \varepsilon \\
    \varepsilon &\sim N(0, 1)
\end{align*}
\end{small}
We designate exam grade the outcome variable and generate the observed outcome $Y_i^{obs}$ for individual $i$ under a binary treatment $z_i \in \{0, 1\}$ and composition index $c(i)$ as 
\begin{align}
    Y_i^{obs} = \text{exam}_i + \mathbf{e}_{c(i)}^T \boldsymbol{\tau} z_i
\end{align}
where $\mathbf{e}_{c(i)}\in \mathbb{R}^{l}$ is a vector that has a 1 in the $c(i)^{th}$ entry and a 0 everywhere else, and $\boldsymbol{\tau} \in \mathbb{R}^{l}$ is a vector of treatment effects. Entry $j$ in $\boldsymbol{\tau}$ can be interpreted as the effect of the intervention for the $j^{th}$ composition type. We run simulations with groups of composition types $\{c_a^1, c_a^2, c_a^3\} = \{0.5, 0.3, 0.7\}$ (that is, 50\% men, 30\% men, and 70\% men), with 20 individuals per group.

\subsection{Step-by-Step Application of IGR}
\subsubsection{Specification}
We specify the target effect estimands as the average treatment effects under particular group compositions of interest. We specify one inspection metric that inspects allocations for balance and one for whether desired group compositions are realized.
For balance, we extend the commonly used Mahalanobis distance metric to the group formation experiment setting. Let $\mathbf{z} \in \{0, 1\}^{N_G}$ denote the vector of treatment assignments across groups. Since there are both random treatment assignments $\mathbf{z}$ and random group assignments $\mathbf{g}$, we use fitness functions of the form $f(\mathbf{z}, \mathbf{g}, \mathbf{X})$.
We define $m^{(\texttt{Mahalanobis-G})}$ as
\begin{align}\label{eq:metric-smd-mahalanobis-g}
    m^{(\texttt{Mahalanobis-G})}(\mathbf{z}, \mathbf{g}, \mathbf{X}) &= 
        \max_{(h_1, h_2) \in \mathcal{G} \times \mathcal{G}} 
        \left( \bar{\mathbf{X}}_{\setminus a}^{(h_1)} - \bar{\mathbf{X}}_{\setminus a}^{(h_2)} \right)
        S_{\setminus a}^{-1}
        \left( \bar{\mathbf{X}}_{\setminus a}^{(h_1)} - \bar{\mathbf{X}}_{\setminus a}^{(h_2)} \right)^T
\end{align}
where $\mathcal{G} = \{0, 1, 2, \cdots, N_G-1\}$ is the set of group indices, $\bar{X}_{\cdot j}^{(g)}$ and $s_{\cdot j}^{(h)}$ are the mean and standard deviation of covariate $j$ in group $h$, $\mathbf{\bar{X}}_{\setminus a}^{(h)}$ is the vector of covariate means in group $h$ excluding the salient attribute $a$, and $S_{\setminus a}$ is the sample covariance matrix excluding $a$. 
To ensure allocations have the desired group compositions, we define the metric $m^{(\texttt{DesiredComps})}$, which takes on value 1 if all desired compositions are realized in at least two groups and is 0 otherwise.
Combining the balance and composition realization metrics, we define the fitness function
\begin{equation}\label{eq:ff-smd+descomps}
\begin{split}
f^{(\texttt{Mahalanobis-GD})}(\mathbf{z}, \mathbf{g}, \mathbf{X})
&= \begin{cases}
    m^{(\texttt{Mahalanobis-G})}(\mathbf{z}, \mathbf{g}, \mathbf{X}) & m^{(\texttt{DesiredComps})}(\mathbf{z}, \mathbf{g}, \mathbf{X}) = 1 \\
    \infty & m^{(\texttt{DesiredComps})}(\mathbf{z}, \mathbf{g}, \mathbf{X}) = 0
\end{cases}
\end{split}
\end{equation}
\subsubsection{Enumeration}
For simplicity, we assume that we want to form groups of equal sizes such that there are exactly two groups with each composition, one of which is assigned treatment and one assigned control. To enumerate a candidate allocation, we first block on the salient attribute, then randomly sample subsets of individuals to place into groups such that the compositions of interest are achieved. For each pair of groups with the same composition, we randomly assign one to treatment and one to control. Under this group formation randomization assignment mechanism, $m^{(\texttt{DesiredComps})}(\mathbf{z}, \mathbf{g}, \mathbf{X})$ is guaranteed to equal 1. We refer to this assignment mechanism as group formation randomization (GFR). Note that, if we were to instead sample candidate allocations from an assignment mechanism that is agnostic to group composition, it would require enumerating an extremely large number of allocations to obtain any allocations that have the desired group compositions. Using the GFR assignment mechanism, we enumerate $M = 100,000$ candidate allocations. 
\subsubsection{Restriction}
We use the restriction rule $r(s) = 0 \text{ if } \ s \geq s^*, r(s) = 1 \text{ if } s < s^* $, where $s^*$ is the $0.5^{\text{th}}$-percentile of scores for the enumerated candidate allocations. This restriction rule filters out all but the top $0.5^{\text{th}}$-percentile of the $100,000$ enumerated allocations, resulting in $m = 500$ accepted allocations.
\subsubsection{Evaluation \& Adaptation}
As with the other simulated experiment, we perform visual checks to evaluate the fitness function and restriction rule. As before, we use a histogram of scores to evaluate the discriminatory power of the fitness function and a histogram of pairwise assignment correlation to evaluate over-restriction. Because we have a single continuous inspection metric, we do not need to evaluate desiderata trade-offs. Instead, we want to ensure that we achieve good balance on each individual covariate, which is not necessarily guaranteed since we used a balance metric that averages across covariates. To check for per-covariate balance, we construct a boxplot of the standardized mean difference between treatment arms for each covariate.
\begin{figure}[ht!]
    \centering
    \includegraphics[width=\linewidth]{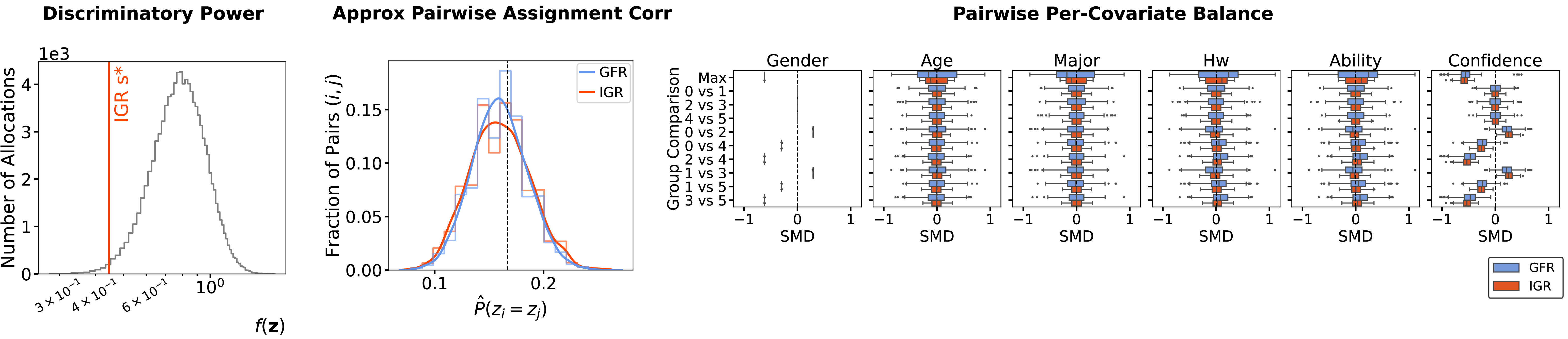}
    \caption{Visual checks for discriminatory power, over-restriction (via approximate pairwise assignment correlation), and balance under fitness function $f^{(\texttt{Mahalanobis-GD})}$. Note that $ability$ and $confidence$ are latent attributes and were not included in the balance metric.}
    \label{fig:enter-label}
\end{figure}

\subsubsection{Pre-registration}
We pre-register the set of 500 accepted allocations, each with equal probability $\frac{1}{500}$ of being drawn. Additionally, we pre-register the fitness function $f^{(\texttt{Mahalanobis-GD})}$ and the threshold restriction rule $r$.

\subsubsection{Randomization} We sample one of the 500 accepted allocations uniformly at random.

\subsection{How IGR Improves Effect Estimates}
In Table \ref{app-tab:groupform-estimates}, we show that, as expected, IGR with both a balance and composition metric reduces the root mean-squared error (RMSE) of treatment effect estimates compared to a benchmark group formation randomization that uses a composition metric alone.
\begin{table}[!ht]
    \centering
    \begin{adjustbox}{width=\textwidth}
    \begin{tabular}{lllllllllll}
    \toprule
     Group Comparison &  & \multicolumn{3}{c}{0 vs 1 $(c_a^1 = 0.5, \text{ effect size } d = 0.3)$} & \multicolumn{3}{c}{2 vs 3 $(c_a^2 = 0.3, \text{ effect size } d = 0.5)$} & \multicolumn{3}{c}{4 vs 5 $(c_a^3 = 0.7, \text{ effect size d } = 0.1)$} \\ 
     \cmidrule(lr){3-5}\cmidrule(lr){6-8}\cmidrule(lr){9-11}
     &  & RMSE & $\% \text{RMSE}_{GFR}$ & Rej. Rate & RMSE & $\% \text{RMSE}_{GFR}$ & Rej. Rate & RMSE & $\% \text{RMSE}_{GFR}$ & Rej. Rate \\
    Design & Fitness &  &  &  &  &  &  &  &  &  \\
    \midrule
    GFR &  & \large{0.39}$\pm$\small{0.02} & Ref & \large{0.16}$\pm$\small{0.02} & \large{0.40}$\pm$\small{0.02} & Ref & \large{0.35}$\pm$\small{0.04} & \large{0.39}$\pm$\small{0.02} & Ref & \large{0.05}$\pm$\small{0.00} \\
    \multirow[t]{2}{*}{IGR} & MaxMahalanobis-GD &  & \large{56.82}$\pm$\small{2.97} & \large{0.38}$\pm$\small{0.04} &  & \large{55.52}$\pm$\small{4.41} & \large{0.80}$\pm$\small{0.04} &  & \large{56.39}$\pm$\small{3.69} & \large{0.08}$\pm$\small{0.01} \\
     & SumMaxAbsSMD-GD &  & \large{48.83}$\pm$\small{2.69} & \large{0.46}$\pm$\small{0.03} &  & \large{47.61}$\pm$\small{3.91} & \large{0.90}$\pm$\small{0.04} &  & \large{47.82}$\pm$\small{3.03} & \large{0.08}$\pm$\small{0.01} \\
    \bottomrule
    \end{tabular}
    \end{adjustbox}
    \caption{\textbf{Root mean-squared error (RMSE) in treatment effect estimates and rejection rate of sharp null for IGR in a group formation experiment}. IGR with both a balance and a composition realization metric reduces RMSE across group comparisons compared to a benchmark group formation randomization (GFR; equivalently, IGR with a composition metric alone). Means and standard deviations are computed over 5 data samples.}
    \label{app-tab:groupform-estimates}
\end{table}

\clearpage
\section{IGR Bias Properties}\label{app-sec:proofs}
In this section, we prove bias properties in the difference-in-means estimator applied to an IGR-designed experiment.

\subsection{Unbiasedness Through the Mirror Property}\label{app-subsec:proofs-unbiased-mirror}
\begin{assumption}[Equal-sized Arms]
    We assume that the design assigns the same number of individuals to the treated and control arms, so $\sum_{i=1}^{n}Z_i = \sum_{i=1}^{n}1 - Z_i = \frac{n}{2}$.
\end{assumption}\label{assump:eq-arm}
\begin{proof}[Proof of Proposition \ref{prop:bias}]
    The proof is identical to that of \cite{Morgan2012-mb} for unbiasedness in re-randomization. Because $\tilde{P}_{IGR}$ satisfies the mirror property, by definition, $\tilde{P}_{IGR}(\mathbf{Z} = \mathbf{z}) = \tilde{P}_{IGR}(\mathbf{Z} = \mathbf{1} - \mathbf{z})$. By implication, $\tilde{P}_{IGR}(Z_i = 0) = \tilde{P}_{IGR}(Z_i = 1) = \frac{1}{2}$. The difference-in-means estimator is $\hat{\tau} := \frac{\sum_{i=1}^n Y_iZ_i}{\sum_{i=1}^n Z_i} - \frac{\sum_{i=1}^n Y_i(1-Z_i)}{\sum_{i=1}^n 1 - Z_i} = \frac{\sum_{i=1}^n Y_iZ_i}{n/2} - \frac{\sum_{i=1}^n Y_i(1-Z_i)}{n/2}$, where the second equality holds under Assumption \ref{assump:eq-arm}. Then we have, 
    \begin{align*}
        E_{\mathbf{Z} \sim \tilde{P}_{IGR}}[\hat{\tau}] 
        &= \frac{\sum_{i=1}^n E[Y_iZ_i]}{n/2} - \frac{E[Y_i(1-Z_i)]}{n/2} \\
        &= \frac{\sum_{i=1}^n E[Y_iZ_i \mid Z_i = 1]\tilde{P}_{IGR}(Z_i = 1) + \sum_{i=1}^n E[Y_iZ_i \mid Z_i = 0]\tilde{P}_{IGR}(Z_i = 0)}{n/2} - \\
        &\quad \frac{\sum_{i=1}^n E[Y_i(Z_i) \mid Z_i = 1]\tilde{P}_{IGR}(Z_i = 1) + \sum_{i=1}^n E[Y_i(1 - Z_i) \mid Z_i = 0]\tilde{P}_{IGR}(Z_i = 0)}{n/2} \\
        &= \frac{\sum_{i=1}^nY_i(1) \cdot \frac{1}{2}}{n/2} - \frac{\sum_{i=1}^nY_i(0) \cdot \frac{1}{2}}{n/2} \quad\quad \text{ by consistency and mirror property} \\
        &= \frac{1}{n}\sum_{i=1}^nY_i(1) - Y_i(0)
    \end{align*} 
\end{proof}
\subsection{Bias in the Presence of Interference}\label{app-subsec:proofs-bias-interference}
When interference is present, a common effect estimand of interest is the global average treatment effect, $\tau_{GATE} = \frac{1}{n}\sum_{i=1}^n Y_i(\mathbf{1}) - Y_i(\mathbf{0})$. In general, designs that satisfy the mirror property and have equal-sized arms do not produce unbiased difference-in-means estimates of the global average treatment effect. Thus, while violation of the mirror property can introduce some bias, it may be ``worth it", so long as it sufficiently reduces interference bias.
\subsubsection{Bias of an Unrestricted Design}
Let $Y_i(Z_i, \mathbf{Z}_{-i})$ denote the potential outcome under the assignment vector where individual $i$ is assigned to $Z_i$ and the remaining individuals to $\mathbf{Z}_{-i}.$
Consider an unrestricted design that satisfies the mirror property. Since $P(\mathbf{Z} = \mathbf{z}) = P(\mathbf{Z} = 1- \mathbf{z})$, it also holds that $P(Z_i = 1) = P(1 - Z_i = 1) = \frac{1}{2}$. Then, under Assumption \ref{assump:eq-arm} we have, 
    \begin{align*}
        E[\hat{\tau}] &= \sum_{i=1}^n \frac{E[Y_i Z_i]}{\frac{n}{2}} - \frac{E[Y_i (1 - Z_i)]}{\frac{n}{2}}  \\ 
        &= \sum_{i=1}^n \frac{E[Y_i(1, \mathbf{Z}_{-i})] \cdot P(Z_i = 1)}{\frac{n}{2}} - \frac{E[Y_i(0, \mathbf{Z}_{-i})]\cdot P(Z_i = 0)}{\frac{n}{2}} \\
        &= \frac{1}{n}\sum_{i=1}^{n}E[Y_i(1, \mathbf{Z}_{-i})] - E[Y_i(0, \mathbf{Z}_{-i})]
    \end{align*}
where the expected value is taken over the possible assignment vectors sampled from the design. The bias in the estimate is therefore
    \begin{align}
        \text{Bias}_{\text{unrestricted}} 
        &= E[\hat{\tau}] - \tau_{GATE} \nonumber \\
        &= \frac{1}{n}\sum_{i=1}^{n}E[Y_i(1, \mathbf{Z}_{-i})] - E[Y_i(0, \mathbf{Z}_{-i})] - \frac{1}{n}\sum_{i=1}^{n}Y_i(\mathbf{1}) - Y_i(\mathbf{0})
    \end{align}\label{eq:bias-unrestricted}
\subsubsection{Bias of an IGR Design}\label{sec:bias-igr}
Now, consider instead an IGR design with fitness function $f$ and restriction rule $r$, such that an assignment vector $\mathbf{z}$ is accepted if and only if $r(f(\mathbf{z})) = 1$.  Let $\mathcal{A}$ denote the event $r(f(\mathbf{z})) = 1$. Suppose this design does not satisfy the mirror property, so $P(\mathbf{Z} = \mathbf{z}) \neq P(\mathbf{Z} = 1 - \mathbf{z})$ and therefore $P(Z_i = 1 \mid \mathcal{A})$ may not be equal to $\frac{1}{2}$. Instead, for every $i$, there exists some perturbation $\Delta_i \in (-\frac{1}{2}, \frac{1}{2})$ such that $P(Z_i = 1 \mid \mathcal{A}) = \frac{1}{2} + \Delta_i$ and  $P(Z_i = 0 \mid \mathcal{A}) = \frac{1}{2} - \Delta_i$. We continue to operate under Assumption \ref{assump:eq-arm}, so that the treatment and control arms are equally sized.
\begin{proposition}[Zero-sum perturbation]
    Under Assumption \ref{assump:eq-arm}, $\sum_{i=1}^n \Delta_i = 0$.
\end{proposition}\label{prop:zero-sum-perturb}
\begin{proof}
    Since $Z_i \in \{0, 1\}$, we have $E[Z_i \mid \mathcal{A}] = P(Z_i = 1 \mid \mathcal{A}) = \frac{1}{2} + \Delta_i$. Thus, 
    \begin{align*}
        \sum_{i=1}^n E[Z_i \mid \mathcal{A}] 
            &= \sum_{i=1}^n \frac{1}{2} + \Delta_i \\ 
            &= \frac{n}{2} + \sum_{i=1}^n \Delta_i 
    \end{align*}
    Pulling the expectation out of the sum, the left-hand side becomes $\sum_{i=1}^n E[Z_i \mid \mathcal{A}] = E[\sum_{i=1}^n Z_i \mid \mathcal{A}] = \frac{n}{2}$ under Assumption \ref{assump:eq-arm}. Thus, $\sum_{i=1}^n \Delta_i = 0$.
\end{proof}
Recall that in IGR, we enumerate a pool of candidate assignment vectors $\mathcal{Z}_{pool}$, then filter out those that do not meet design desiderata. We can write the difference-in-means estimate under an IGR design as $E_{\mathcal{Z}_{pool}}[\hat{\tau} \mid \mathcal{A}]$. Then, under Assumption \ref{assump:eq-arm} we have,
    \begin{align*}
        E[\hat{\tau} \mid \mathcal{A}] 
        &= 
            \sum_{i=1}^n \frac{E[Y_i Z_i \mid \mathcal{A}]}{\frac{n}{2}} - \frac{E[Y_i (1 - Z_i) \mid \mathcal{A}]}{\frac{n}{2}}  \\ 
        &= \sum_{i=1}^n \frac{E[Y_i(1, \mathbf{Z}_{-i}) \mid \mathcal{A}] \cdot P(Z_i = 1 \mid \mathcal{A})}{\frac{n}{2}} - \frac{E[Y_i(0, \mathbf{Z}_{-i}) \mid \mathcal{A}]\cdot P(Z_i = 0 \mid \mathcal{A})}{\frac{n}{2}} \\
        &= \sum_{i=1}^n \frac{E[Y_i(1, \mathbf{Z}_{-i}) \mid \mathcal{A}] \cdot (\frac{1}{2} + \Delta_i)}{\frac{n}{2}} - \frac{E[Y_i(0, \mathbf{Z}_{-i}) \mid \mathcal{A}]\cdot (\frac{1}{2} - \Delta_i)}{\frac{n}{2}} \\
        &= \frac{1}{n}\sum_{i=1}^n E[Y_i(1, \mathbf{Z}_{-i}) \mid \mathcal{A}] - E[Y_i(0, \mathbf{Z}_{-i}) \mid \mathcal{A}] +
            \frac{2}{n}\sum_{i=1}^n \Delta_i \left(E[Y_i(1, \mathbf{Z}_{-i}) \mid \mathcal{A}] + E[Y_i(0, \mathbf{Z}_{-i}) \mid \mathcal{A}] \right)
    \end{align*}
The bias in the estimate is therefore
    \begin{align}
        \text{Bias}_{\text{IGR}}
        &= E[\hat{\tau} \mid \mathcal{A}] - \tau_{GATE} \nonumber \\ 
        &= \frac{1}{n}\sum_{i=1}^n E[Y_i(1, \mathbf{Z}_{-i}) \mid \mathcal{A}] - E[Y_i(0, \mathbf{Z}_{-i}) \mid \mathcal{A}] \nonumber \\
            & \quad\quad + \frac{2}{n}\sum_{i=1}^n \Delta_i \left(E[Y_i(1, \mathbf{Z}_{-i}) \mid \mathcal{A}] + E[Y_i(0, \mathbf{Z}_{-i}) \mid \mathcal{A}] \right) \nonumber \\
            & \quad\quad - \frac{1}{n}\sum_{i=1}^{n}Y_i(\mathbf{1}) - Y_i(\mathbf{0})
    \end{align}\label{eq:bias-igr}
IGR reduces absolute bias compared to an unrestricted design if $|\text{Bias}_{\text{IGR}}| <  |\text{Bias}_{\text{unrestricted}}|$. 
\subsubsection{Bias of an Unrestricted vs IGR design for an Additive Outcome Model}\label{subsec:bias-additive-mdl}
When does the above inequality hold? To gain some insight, we look at a specific outcome model with an additive direct effect and an additive spillover effect on individuals assigned to the control arm.
\begin{definition}[Additive outcome model]
    \begin{align}
        Y_i(\mathbf{Z}) &= \alpha_i + \tau Z_i + \delta h(\mathbf{Z}_{-i})(1 - Z_i)
    \end{align}
    where $\alpha_i$, $\tau$, $\delta$ are scalars and $h: \mathbb{R}^N \times \mathbb{R}^{N \times d} \mapsto \mathbb{R}_{\geq 0}$ is an exposure function that maps the treatment assignments of everyone other than $i$ to a non-negative scalar value. Assume $h(\mathbf{0}) = 0$.
\end{definition}
With this outcome model, we have $\tau_{GATE} = \frac{1}{n}\sum_{i=1}^n \alpha_i + \tau - \alpha_i = \tau$. We also have $Y_i(1, \mathbf{Z}_{-i}) = \alpha_i + \tau$ and $Y_i(0, \mathbf{Z}_{-i}) =\alpha_i + \delta h(\mathbf{Z}_{-i})$. Thus, we can write $\text{Bias}_{\text{Unrestricted}}$ as
\begin{align*}
    \text{Bias}_{\text{Unrestricted}}
    &= \frac{1}{n}\sum_{i=1}^nE[\alpha_i + \tau] - E[\alpha_i + \delta h(\mathbf{Z_{-i}})] - \tau \\
    &= \frac{1}{n}\sum_{i=1}^{n} -\delta E[ h(\mathbf{Z}_{-i})]
\end{align*}
We can write $\text{Bias}_{\text{IGR}}$ as
\begin{align*}
    \text{Bias}_{\text{IGR}} 
    &= \frac{1}{n}\sum_{i=1}^n E[\alpha_i + \tau \mid \mathcal{A}] - E[\alpha_i + \delta h(\mathbf{Z}_{-i}) \mid \mathcal{A}] \\
    & \quad\quad + \frac{2}{n}\sum_{i=1}^n\Delta_i \left( E[\alpha_i + \tau \mid \mathcal{A}] + E[\alpha_i+ \delta h(\mathbf{Z}_{-i}) \mid \mathcal{A}] \right) \\
    & \quad\quad - \tau \\
    &= \frac{1}{n} \sum_{i=1}^n -E[\delta h(\mathbf{Z}_{-i}) \mid \mathcal{A}] + \frac{2}{n}\sum_{i=1}^n\Delta_i E[\delta h(\mathbf{Z}_{-i}) \mid \mathcal{A}] + \frac{4}{n}\sum_{i=1}^n \Delta_i \alpha_i\\
        & \quad\quad \text{by Proposition \ref{prop:zero-sum-perturb}} \\
    &= \frac{1}{n}\sum_{i=1}^n -\delta (1 - 2\Delta_i) E[h(\mathbf{Z}_{-i}) \mid \mathcal{A}] + \frac{4}{n}\sum_{i=1}^n \Delta_i \alpha_i
\end{align*}
Without loss of generality, we assume $\delta > 0$ and $\sum_{i=1}^n \alpha_i = 0$. Note that since both $\delta > 0$ and $h(\mathbf{Z}_{-i}) \geq 0$ for all $\mathbf{Z}$, this implies $\text{Bias}_{\text{Unrestricted}} \leq 0$. We will focus on the case where $\text{Bias}_{\text{IGR}} \leq 0$, i.e. where IGR does not change the sign of the bias term.
\begin{align*}
     &\left|\text{Bias}_{\text{IGR}}| < |\text{Bias}_{\text{Unrestricted}}\right| \\
     \Leftrightarrow & \text{Bias}_{\text{IGR}}  >  \text{Bias}_{\text{Unrestricted}} \quad\quad \text{ since biases assumed non-positive} \\
    \Leftrightarrow & \frac{1}{n}\sum_{i=1}^n \delta (1-2\Delta_i)E[h(\mathbf{Z}_{-i}) \mid \mathcal{A}] - \frac{4}{n}\sum_{i=1}^n \Delta_i \alpha_i < \frac{1}{n}\sum_{i=1}^n \delta E[h(\mathbf{Z}_{-i})] \\
    \Leftrightarrow & \frac{1}{n}\sum_{i=1}^n (1-2\Delta_i)E[h(\mathbf{Z}_{-i}) \mid \mathcal{A}] - \frac{4}{\delta n}\sum_{i=1}^n \Delta_i \alpha_i < \frac{1}{n}\sum_{i=1}^n E[h(\mathbf{Z}_{-i})] \\
     \Leftrightarrow &
     \underbrace{\frac{\sum_{i=1}^n (1-2\Delta_i)E[h(\mathbf{Z}_{-i}) \mid \mathcal{A}]}{\sum_{i=1}^n (1 - 2 \Delta_i)}}_{\text{weighted avg of expected exposure}}
     - \underbrace{\frac{4}{\delta n}\sum_{i=1}^n \Delta_i \alpha_i}_{\text{perturbation balance}}
     < \underbrace{\frac{1}{n}\sum_{i=1}^n E[h(\mathbf{Z}_{-i})]}_{\text{avg of expected exposure}} & \text{by Proposition \ref{prop:zero-sum-perturb}}
\end{align*}
IGR reduces absolute bias compared to an unrestricted design if the above inequality holds. The left-hand side of the inequality is the difference of two terms: the weighted average of the expected exposure under the restricted design and the scaled sample covariance of $\Delta$ and $\alpha$. The right-hand side of the inequality is the expected exposure under the unrestricted design. 

First, we dissect the term for the weighted average of the expected exposure under the restricted design. For $i$ such that $\Delta_i \in (-\frac{1}{2}, 0)$ (i.e., $i$ is assigned to control with probability greater than $\frac{1}{2}$), the weight satisfies $(1 - 2\Delta_i) \in (1, 2)$. For $i$ such that $\Delta_i \in [0, \frac{1}{2})$ (i.e., $i$ is assigned to treatment with probability greater than $\frac{1}{2}$), the weight satisfies $(1 - 2\Delta_i) \in (0, 1)$. From this, we can see that it is ``worth it" to assign some individuals to control with probability greater than $\frac{1}{2}$ -- thus applying greater weight to their expected exposure in the summation -- if in doing so, we are able to substantively reduce their expected exposure, $E[h(\mathbf{Z}_{-i}) \mid \mathcal{A}]$.

At the same time, we must account for the scaled sample covariance of $\Delta$ and $\alpha$. If we want to keep the left-hand side of the inequality small, then we want this covariance term to be positive (or negative, but close to zero). In other words, we do not want to systematically set the perturbation $\Delta_i \in (-\frac{1}{2}, 0)$ (preferential assignment to control) when $\alpha_i > 0$ or $\Delta_i \in (0, \frac{1}{2})$ (preferential assignment to treatment) when $\alpha_i < 0$.

\clearpage
\section{Why Pre-register: Type I Error Rate Inflation in Re-randomization}\label{app-sec:phack-rerand}
Pre-registration is included explicitly in IGR and is absent from prior similar restricted randomization approaches, e.g. re-randomization. In this section, we illustrate the importance of pre-registering the accepted allocations in a restricted design and using exactly these pre-registered allocations to do randomization inference in the analysis phase. Specifically, we show that we can easily inflate Type I error in re-randomization by using one balance threshold to re-randomize in the design phase and a different balance threshold to sample allocations for the randomization test in the analysis phase. Much of the results from this section rely on heavy lifting done by \cite{Krieger2020-hl} and  \cite{Li2018-gh}.

\subsection{Preliminary Set-up and Notation}
Suppose we are running a two-arm experiment on $n$ study units with observed covariates $\mathbf{X} \in \mathbb{R}^{n \times k}$. Let $\mathbf{Z} \in \{0, 1\}^n$ denote the random treatment assignment vector variable. We denote the potential outcomes under treatment and control as $Y_i(1)$ and $Y_i(0)$, respectively. The individual level causal effect is defined as $\tau_i := Y_i(1) - Y_i(0)$, and the sample average treatment effect is our target estimand, $\tau := \frac{\sum_{i=1}^N \tau_i}{n}$. 
We assume that our design is restricted to only include assignment vectors where the proportion of units assigned to treatment is fixed, $p_T = \frac{1}{n}\sum_{i=1}^NZ_i$. We then further restrict the design through rerandomization. To sample the observed assignment vector for the experiment, we apply rerandomization with Mahalanobis distance as our balance metric. For assignment vector $\mathbf{z}$, the corresponding Mahalanobis distance is
\begin{align*}
    M &:= (\bar{\mathbf{X}}_T -  \bar{\mathbf{X}}_C)^T 
    \text{Cov}(\bar{\mathbf{X}}_T -  \bar{\mathbf{X}}_C)^{-1}
    (\bar{\mathbf{X}}_T -  \bar{\mathbf{X}}_C) \\
    &= \sqrt{n}(\bar{\mathbf{X}}_T -  \bar{\mathbf{X}}_C)^T(p_Tp_C\text{Cov}(\mathbf{X}))^{-1}\sqrt{N}(\bar{\mathbf{X}}_T -  \bar{\mathbf{X}}_C)
\end{align*}
where $\bar{\mathbf{X}}_T$ is the mean covariate vector for the treated units and $\bar{\mathbf{X}}_C$ for the control units. In order to encourage covariate balance, a treatment assignment vector $\mathbf{z}$ is accepted if and only if $M \leq a$.
To estimate the average treatment effect, we use the standard difference-in-means estimator, $\hat{\tau} := \frac{\sum_{i=1}^nY_iZ_i}{p_Tn} - \frac{\sum_{i=1}^nY_i(1 - Z_i)}{p_Cn}$. We use a randomization test to test the one-sided null hypothesis, $H_0: \tau \leq 0$ with alternative hypothesis $H_a: \tau > 0$. Let $\mathbf{z}_{obs}$ denote the observed assignment vector that was used to run the experiment and $\hat{\tau}_{obs}$ denote the difference-in-means estimate under $\mathbf{z}_{obs}$. To form an empirical null distribution for $\hat{\tau}_{obs}$, we compute the difference-in-means over a set of $2D$ assignment vectors, $\{\mathbf{z}_1, \mathbf{z}_1^m, \mathbf{z}_2, \mathbf{z}_2^m, \dots, \mathbf{z}_{D}, \mathbf{z}_{D}^m\}$, where $\mathbf{z}_j^m = \mathbf{1} - \mathbf{z}_j$ is $\mathbf{z}_j$'s mirror assignment vector. We include mirror assignment vectors so that the difference-in-means estimate is unbiased. We define $\hat{\tau}_j$,  $\hat{\tau}^m_j$ as 
\begin{align*}
    \hat{\tau}_j &= \frac{\sum_{i=1}^nY_i^{obs}z_{ji}}{p_T \cdot n} - \frac{\sum_{i=1}^nY_i^{obs}(1 - z_{ji})}{p_C \cdot n} \\
    \hat{\tau}_j^m &= -\hat{\tau}_j
\end{align*}

From \cite{Krieger2020-hl}, the approximate rejection rate for a one-sided level $\alpha$ test is
\begin{align}\label{eq:rej-rate}
    \mathcal{R} = 
    P \left (
        \hat{\tau}_{obs} > \text{Quantile}
            [\{\hat{\tau}_{1}, \hat{\tau}_{1}^m, \dots \hat{\tau}_{D}, \hat{\tau}_{D}^m\}, 1- \alpha]
    \right)
\end{align}
where we define the quantile function for some $p \in [0, 1]$ and $\mathcal{W} = \{ w_1, w_2, \dots, w_M\}$ as
\begin{align}\label{eq:quantile}
    \text{Quantile}\left[ 
        \mathcal{W}; p
    \right] := 
    \min 
    \{
        w \in \mathcal{W} \mid \sum_{i=1}^M \mathbf{1}\{ w_i < w\} > pM
    \}
\end{align}

Notice that the assignment vectors that are used in the randomization test are sampled during the analysis phase. We show that, when the true effect is zero, we can inflate the rejection rate if we sample these vectors with a balance criterion that is different from the one used in the design phase. We let $M \leq a$ be the acceptance criterion in the design phase. Similarly, we let $M \leq a', \ a' \neq a$ be the acceptance criterion in the analysis phase, so that the set of assignment vectors used to build the empirical null distribution all must satisfy $M \leq a'$. We use $\mathcal{M}_{a}$ to denote the event that a vector $\mathbf{z}$ is accepted under the design phase acceptance criterion and $\mathcal{M}_{a'}$ to denote the event that a vector $\mathbf{z}$ is accepted under the analysis phase acceptance criterion.

\subsection{Asymptotic distributions}
We first introduce the asymptotic distribution derived in \cite{Li2018-gh}. Let $S^2_{Y(z)} = \sum_{i=1}^N Y_i(z) - \bar{Y}(z) / (n-1)$ denote the finite population variance of the potential outcomes and $S_{\tau}^2 = (\sum_{i=1}^N (\tau_i - \tau)^2) / (n-1)$ be the finite population variance of the individual causal effects. Let $V_{\tau\tau} = (n / n_T)S_{Y(1)}^2 + (n / n_C)S_{Y(0)}^2 - S^2_{\tau}$. 

Under Condition 1 and Theorem 1 in \cite{Li2018-gh}, the asymptotic sampling distribution of $\hat{\tau}_{obs}$ can be expressed as a linear combination of two independent random variables: $\varepsilon_0 \sim \mathcal{N}(0, 1)$ and $L_{k,a} \sim \chi_{k, a} S \sqrt{\beta_k}$, where $S$ is a random sign taking $\pm 1$ with probability $1/2$ and $\beta_k \sim \text{Beta}(1/2, (k-1)/2)$. Specifically, the asymptotic distribution for the observed estimate $\hat{\tau}_{obs}$ is
\begin{align}\label{eq:asymp-tau-obs}
    \sqrt{n}(\hat{\tau}_{obs} - \tau) \mid \mathcal{M}_a \ \dot{\sim} \sqrt{V_{\tau\tau}}\left( \sqrt{1 - R^2} \cdot \varepsilon_0 + \sqrt{R^2} \cdot L_{k, a} \right)
\end{align}

where $R^2$ is a measure of association between the covariates and the potential outcomes as defined in Proposition 1 of \cite{Li2018-gh}, and $\dot{\sim}$ is the authors' notation for two sequences of random variables converging weakly to the same distribution. Since we want to illustrate how we can inflate the rejection rate, we assume that the ground truth individual-level causal effects are all zero. That is, $\tau_i = Y_i(1) - Y_i(0) = 0$ for all $i$, which also means $\tau = \frac{1}{N}\sum_{i=1}^{n} \tau_i = 0$. Under this assumption, the asymptotic sampling distribution for all  $\hat{\tau}_j$ can be written in the same way as for $\hat{\tau}_{obs}$, replacing $a$ with $a'$
\begin{align}\label{eq:asymp-tau}
    \sqrt{n}(\hat{\tau}_{j} - \tau) \mid \mathcal{M}_{a'} \ \dot{\sim} \sqrt{V_{\tau\tau}}\left( \sqrt{1 - R^2} \cdot \varepsilon_0 + \sqrt{R^2} \cdot L_{k, a'} \right)
\end{align}
Since $\hat{\tau}_j^m = -\hat{\tau}_j$, we also have 
\begin{align}\label{eq:asymp-tau-mirror}
    \sqrt{n}(\hat{\tau}_{j}^m - \tau) \mid \mathcal{M}_{a'} \ \dot{\sim} -\sqrt{V_{\tau\tau}}\left( \sqrt{1 - R^2} \cdot \varepsilon_0 + \sqrt{R^2} \cdot L_{k, a'} \right)
\end{align}

\subsection{Rejection rate in the limit}
 In this section, we derive an expression for the rejection rate in the limit as $n$ and $D$ go to infinity. We first establish notation for known limiting distributions. Let $\tau_{obs, \infty}$ denote the limiting distribution for $\hat{\tau}_{obs}$. By \ref{eq:asymp-tau-obs}, we have 
\begin{align*}
    \sqrt{n}(\hat{\tau}_{obs} - \tau) \mid \mathcal{M}_a &\xrightarrow{d} \tau_{obs, \infty} \\
    \sqrt{V_{\tau\tau}} \left( \sqrt{1 - R^2} \cdot \varepsilon_0 + \sqrt{R^2} \cdot L_{k,a} \right) &\xrightarrow{d} \tau_{obs, \infty}
\end{align*}
Let $V_{\tau\tau, \infty} := \lim_{n \rightarrow \infty} V_{\tau\tau}$ and $R^2_{\infty} := \lim_{n \rightarrow \infty} R^2$ denote the limits of $V_{\tau\tau}$ and $R^2$. We assume that these limits exist and are finite. Then we can write $\tau_{obs, \infty}$ as 
\begin{align}
    \tau_{obs, \infty} \overset{d}{=} \sqrt{V_{\tau\tau, \infty}} \left( \sqrt{1 - R^2_{\infty}} \cdot \varepsilon_0 + \sqrt{R^2_{\infty}} \cdot L_{k,a} \right)
\end{align}
where $\overset{d}{=}$ indicates equivalence in distribution. 

All $\hat{\tau}_j$ are independent and identically distributed, and all $\hat{\tau}^m_{j}$ are independent and identically distributed. Thus, all $\hat{\tau}_j$ have the same limiting distribution, and all $\hat{\tau}^m_{j}$ have the same limiting distribution. Let $\tau_{\infty}$ denote the limiting distribution for $\hat{\tau}_j$ and $\tau_{\infty}^m$ denote the limiting distribution for $\hat{\tau}_j^m$. By the asymptotics established in \ref{eq:asymp-tau} and \ref{eq:asymp-tau-mirror},
\begin{align}
    \tau_{\infty} &\overset{d}{=}  \sqrt{V_{\tau\tau, \infty}} \left( \sqrt{1 - R^2_{\infty}} \cdot \varepsilon_0 + \sqrt{R^2_{\infty}} \cdot L_{k,a'} \right) \\  
    \tau_{\infty}^m &\overset{d}{=}  -\sqrt{V_{\tau\tau, \infty}} \left( \sqrt{1 - R^2_{\infty}} \cdot \varepsilon_0 + \sqrt{R^2_{\infty}} \cdot L_{k,a'} \right)
\end{align}
We define $\hat{\tau}^+_{j}:= \max (\hat{\tau}_j, \hat{\tau}^m_j) = |\hat{\tau}_j|$, where the equality holds because $\hat{\tau}_j = -\hat{\tau}^m_j$. All $\hat{\tau}^+_j$ are independent, since for all $j \neq j'$,   $\hat{\tau}_j$ is independent of both $\hat{\tau}_j'$ and $\hat{\tau}_{j'}^m$. The $\hat{\tau}^+_j$ are also identifically distributed. By the continuous mapping theorem, 
\begin{align}\label{eq:tau-pos-asymp}
    \hat{\tau_j}^+ \rightarrow \tau^+_{\infty} &\overset{d}{=} 
        \left | 
            \sqrt{V_{\tau\tau, \infty}} \left( \sqrt{1 - R^2_{\infty}} \cdot \varepsilon_0 + \sqrt{R^2_{\infty}} \cdot L_{k,a'} \right)
        \right |
\end{align}
The approximate rejection rate $\mathcal{R}$ defined in Equation \ref{eq:rej-rate} includes the term $\text{Quantile}[\{\hat{\tau}_j, \hat{\tau}_j^m\}_{j=1}^D; 1-\alpha]$. Below, we show that this term is equivalent to $\text{Quantile}[\{\hat{\tau}_j^+\}_{j=1}^D; 1-2\alpha]$.
\begin{proposition}\label{eq:quantile-pos}
    Let $\mathcal{W} = \{w_1, -w_1, \dots, w_D, -w_D\}$. Define $w^+:= \max{(w_1, -w_1)}$ and $\mathcal{W}^+ = \{w_1^+, \dots, w_D^+\}$. Then for all $\alpha < 0.5$,
    \begin{align*}
        \text{Quantile}[\mathcal{W}; 1 - \alpha] = \text{Quantile}[\mathcal{W}^+; 1 - 2\alpha]
    \end{align*}
\end{proposition}
\begin{proof}
    By definition, we can write 
    \begin{align*}
        \text{Quantile}[\mathcal{W}; 1 - \alpha] = 
        \min \{w \in \mathcal{W} \mid \sum_{i=1}^D \mathbf{1}\{w_i < w\} + \mathbf{1}\{-w_i < w\} > (1 - \alpha)\cdot 2D\}
    \end{align*}
    Let $q := \text{Quantile}[\mathcal{W}; 1 - \alpha], \ q\in \mathcal{W}$. We first observe that $q > 0$ when $\alpha < 0.5$. For sake of contradiction, suppose $q \leq 0$. Since $q$ is the 1 - $\alpha$ quantile and $\alpha < 0.5$, there must exist at least $\lceil(1 - \alpha) \cdot 2D \rceil > D$ elements of $\mathcal{W}$ that are less that $q$. However, we know that there are at least $D$ elements of $\mathcal{W}$ that are non-negative, so there can be no more than $D$ elements of $W$ that are strictly less than $q$. Since $q > 0$, by definition, $q \in \mathcal{W}^+$. Thus, we can rewrite the quantile as 
        \begin{align*}
        \text{Quantile}[\mathcal{W}; 1 - \alpha] = 
        \min \{w \in \mathcal{W}^+ \mid \sum_{i=1}^D \mathbf{1}\{w_i < w\} + \mathbf{1}\{-w_i < w\} > (1 - \alpha)\cdot 2D\}
    \end{align*}
    
    Now, take the term $\mathbf{1}\{w_i < w \} + \mathbf{1}\{-w_i < w\}$ in the summation. We can rewrite this term as $\mathbf{1}\{\min(w_i, -w_i) < w \} + \mathbf{1}\{\max(w_i, -w_i) < w\}$. If $w$ is the $1 - \alpha$ quantile and $\alpha < 0.5$, then by the argument in the previous paragraph, $w > 0$. Thus, we know that $\min(w_i, -w_i) \leq 0 < w$, so $\mathbf{1}\{\min(w_i, -w_i) < w\} = 1$ for all $i$. Thus, $\mathbf{1}\{w_i < w \} + \mathbf{1}\{-w_i < w\} = \mathbf{1}\{\max(w_i, -w_i) < w\} + 1 = \mathbf{1}\{w_i^+ < w\} + 1$.

    Plugging this back into the overall quantile function, we have
    \begin{align*}
        \text{Quantile}[\mathcal{W}; 1 - \alpha] &= 
        \min\{w \in \mathcal{W}^+ \mid \sum_{i=1}^D \mathbf{1} \{w_i^+ < w\} + 1 > (1- \alpha) \cdot 2D\} \\
        &= 
        \min\{w \in \mathcal{W}^+ \mid \sum_{i=1}^D \mathbf{1} \{w_i^+ < w\}  > (1- \alpha) \cdot 2D - D\} \\
        &= \min\{w \in \mathcal{W}^+ \mid \sum_{i=1}^D \mathbf{1} \{w_i^+ < w\}  > (1- 2\alpha) \cdot D\} \\
        &= \text{Quantile}[\mathcal{W}^+; 1- 2\alpha] 
    \end{align*}
\end{proof}
We now introduce the expression for the limiting approximate rejection rate. 

\begin{theorem}\label{thm:lim-rr}
The limiting $\alpha$-level approximate rejection rate $\mathcal{R}$ for any $\alpha < 0.5$ is
\begin{align}
    \lim_{n \rightarrow \infty} \lim_{D \rightarrow \infty} \mathcal{R} = 
    \begin{cases}
        0 & \hat{\tau}_{obs} < 0 \\
        P(\tau_{obs, \infty} > \text{Quantile} \left[\tau^+_{\infty}; 1- 2\alpha \right]) & \text{otherwise}
    \end{cases}
\end{align}
\end{theorem}

\begin{proof}
We want to express the rejection rate $\mathcal{R}$ defined in Equation \ref{eq:rej-rate} in terms of the asymptotic distributions $\tau_{obs, \infty}$ and $\tau_{\infty}^+$. In the proof for Proposition \ref{eq:quantile-pos}, we showed that the $1 - \alpha$ quantile of $\{\hat{\tau}_1, \hat{\tau}_1^m, \dots, \hat{\tau}_D, \hat{\tau}_D^m\}$ must be positive. Thus, $\mathcal{R} = 0$ if $\hat{\tau}_{obs} < 0$. 

Suppose that $\hat{\tau}_{obs} \geq 0$. By Proposition \ref{eq:quantile-pos}, we have 
\begin{align*}
    \text{Quantile}[\{\hat{\tau}_1, \hat{\tau}_1^m, \dots, \hat{\tau}_D, \hat{\tau}_D^m\}; 1- \alpha] = \text{Quantile}[\{\hat{\tau}_1^+, \dots, \hat{\tau}_D^+\}; 1- 2\alpha]
\end{align*}
Recall that $\{\hat{\tau}_1^+, \cdots, \hat{\tau}_D^+\}$ are independent and identically distributed. We first take the limit in the number of sampled allocations, $2D$.
\begin{align*}
    \lim_{D \rightarrow \infty} \mathcal{R} &= 
    \lim_{D \rightarrow \infty} P \left (
        \hat{\tau}_{obs} > \text{Quantile}
            [\{\hat{\tau}_{1}, \hat{\tau}_{1}^m, \dots \hat{\tau}_{D}, \hat{\tau}_{D}^m\}; 1- \alpha]
    \right) \\
    &= \lim_{D \rightarrow \infty} P(\hat{\tau}_{obs} > \text{Quantile}
    \left[ 
        \{\hat{\tau}^+\}_{j=1}^D; 1 - 2\alpha
    \right]) \\
    &= P(\hat{\tau}_{obs} > \text{Quantile} \left[
        \hat{\tau}_j^+; 1- 2\alpha 
    \right])
\end{align*}
where the last equality holds by application of the Glivenko-Cantelli theorem. 

We now take the limit in the number of study units, $n$.
\begin{align*}
    \lim_{n \rightarrow \infty} P\left(\hat{\tau}_{obs} > \text{Quantile} \left[
        \hat{\tau}_j^+; 1- 2\alpha 
    \right]\right) &= 
    \lim_{n \rightarrow \infty} P\left(\sqrt{n}(\hat{\tau}_{obs} - \tau) > \sqrt{n}(\text{Quantile}[\hat{\tau}^+_j; 1- 2\alpha] - \tau)\right) \\
    &= \lim_{n \rightarrow \infty} P\left(\sqrt{n}(\hat{\tau}_{obs} - \tau) > \text{Quantile}[\sqrt{n}(\hat{\tau}^+_j - \tau); 1 - 2\alpha]\right) \\
    &= P\left(\tau_{obs, \infty} > \text{Quantile}[\tau_{\infty}^+; 1-2\alpha]\right).
\end{align*}
The last equality holds because, for a sequence of cumulative distribution functions, $F_n \rightarrow F$ if and only if $F_n^{-1} \rightarrow F^{-1}$ (Lemma 21.1, \cite{van-der-Vaart2012-uu}). Let $F_{n, \tau^+}$ denote the cumulative distribution function for $\sqrt{n}(\hat{\tau}^+ - \tau)$. Then $F^{-1}_{n, \tau^+}(1 - 2\alpha) = \text{Quantile}\left[ \sqrt{n}(\hat{\tau}_i^+ - \tau); 1- 2\alpha \right]$ by definition. By \ref{eq:tau-pos-asymp}, we know that $F_{n, \tau^+} \rightarrow F_{\tau^+}$, where $F_{\tau^+}$ is the cumulative distribution function for $\tau_\infty^+$. Thus, $F^{-1}_{n, \tau^+}(1 - 2\alpha) \rightarrow F_{\tau^+}^{-1}(1 - 2\alpha) = \text{Quantile}[\tau_\infty^+; 1 - 2\alpha]$.
\end{proof}

\subsection{Simulation}
Using the result from Theorem \ref{thm:lim-rr} and the limiting distributions $\tau_{obs, \infty}$ and $\tau_{\infty}^+$ written in terms of $V_{\tau\tau, \infty}, R_{\infty}^2, \varepsilon_0$, $L_{k, a}$, and $L_{k, a'}$, we can simulate the limiting rejection rate under different relationships between $a$ (the design phase balance threshold) and $a'$ (the analysis phase balance threshold).

\begin{figure}[ht!]
    \centering
    \includegraphics[width=0.8\linewidth]{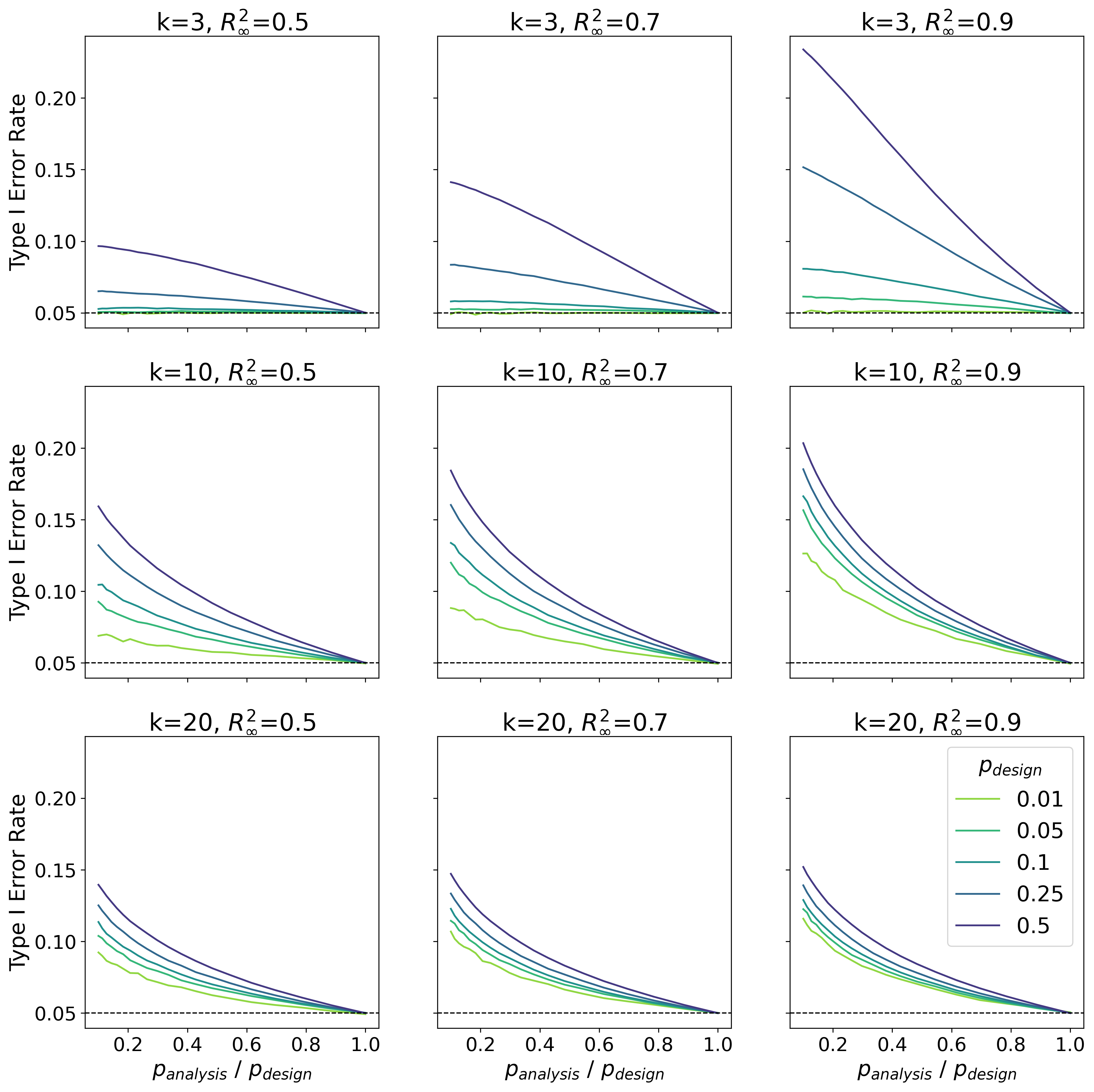}
    \caption{Type I error rate for varying $p_{design}$ and $p_{analysis}$.  The largest amount of Type I error rate inflation occurs when a lenient threshold is used in the design phase (i.e. $p_{design}$  is large) and a strict threshold is used in the analysis phase (i.e. $p_{analysis}$ is small). The greatest inflation occurs when there are few covariates but $R_{\infty}^2$ is large ($k = 3, R_{\infty}^2 = 0.9$).}
    \label{fig:rr}
\end{figure}

We run the simulation with $\alpha = 0.05$, $V_{\tau\tau, \infty} = 1$, $k \in \{3, 10, 20\}$ and $R_{\infty}^2 \in \{0.5, 0.7, 0.9\}$ We vary $a$ such that the probability of an assignment vector being accepted is $p_{design} \in [0.1, 0.9]$. For each $a$, we set $a'$ to values such that the probability of being accepted is $p_{analysis} = cp_{design}$, where $c \in [0.01, 1]$. Figure \ref{fig:rr} shows how different amounts of Type I error inflation can occur depending on the value of $c$, $k$, and $R_{\infty}^2$.

\clearpage
\begin{thebibliography}{}

\bibitem[Aronow and Samii, 2017]{aronow2017-gb}
Aronow, P.~M. and Samii, C. (2017).
\newblock Estimating average causal effects under general interference, with application to a social network experiment.
\newblock {\em Ann. Appl. Stat.}

\bibitem[Athey et~al., 2018]{Athey2018-yf}
Athey, S., Eckles, D., and Imbens, G.~W. (2018).
\newblock Exact p-values for network interference.
\newblock {\em J. Am. Stat. Assoc.}, 113(521):230--240.

\bibitem[Baiocchi et~al., 2017]{Baiocchi2017-om}
Baiocchi, M., Omondi, B., Langat, N., Boothroyd, D.~B., Sinclair, J., Pavia, L., Mulinge, M., Githua, O., Golden, N.~H., and Sarnquist, C. (2017).
\newblock A {Behavior-Based} intervention that prevents sexual assault: the results of a {Matched-Pairs}, {Cluster-Randomized} study in nairobi, kenya.
\newblock {\em Prev. Sci.}, 18(7):818--827.

\bibitem[Baker, 2016]{Baker2016-hy}
Baker, M. (2016).
\newblock 1,500 scientists lift the lid on reproducibility.
\newblock {\em Nature}, 533:452--454.

\bibitem[Basse and Airoldi, 2018]{Basse2018-ii}
Basse, G.~W. and Airoldi, E.~M. (2018).
\newblock Model-assisted design of experiments in the presence of network-correlated outcomes.
\newblock {\em Biometrika}, 105(4):849--858.

\bibitem[Basse et~al., 2024]{Basse2024-sw}
Basse, G.~W., Ding, P., Feller, A., and Toulis, P. (2024).
\newblock Randomization tests for peer effects in group formation experiments.
\newblock {\em Econometrica}, 92(2):567--590.

\bibitem[Box et~al., 2005]{Box2005-we}
Box, G. E.~P., Stuart~Hunter, J., and Hunter, W.~G. (2005).
\newblock {\em Statistics for Experimenters: Design, Innovation, and Discovery}.
\newblock Wiley.

\bibitem[Branson et~al., 2016]{Branson2016-fm}
Branson, Z., Dasgupta, T., and Rubin, D.~B. (2016).
\newblock Improving covariate balance in {2K} factorial designs via rerandomization with an application to a new york city department of education high school study.
\newblock {\em Annals of Applied Statistics}, 10(4):1958--1976.

\bibitem[Cai et~al., 2015]{Cai2015-kn}
Cai, J., De~Janvry, A., and Sadoulet, E. (2015).
\newblock Social networks and the decision to insure.
\newblock {\em Am. Econ. J. Appl. Econ.}, 7(2):81--108.

\bibitem[Carroll et~al., 2023]{Carroll2023-hq}
Carroll, J.~M., Yeager, D.~S., Buontempo, J., Hecht, C., Cimpian, A., Mhatre, P., Muller, C., and Crosnoe, R. (2023).
\newblock Mindset $\times$ context: Schools, classrooms, and the unequal translation of expectations into math achievement.
\newblock {\em Monogr. Soc. Res. Child Dev.}, 88(2):7--109.

\bibitem[Cho et~al., 2021]{Cho2021-pf}
Cho, J., Li, Y., Armstrong, A.~K., Russ, A., Krasny, M.~E., and Kizilcec, R.~F. (2021).
\newblock Using social norms to promote actions beyond the course.
\newblock In {\em Proceedings of the Eighth {ACM} Conference on Learning @ Scale}, L@S '21, pages 161--172. Association for Computing Machinery.

\bibitem[Cho et~al., 2022]{Cho2022-oy}
Cho, J., Li, Y., Kudryavtsev, A., Kizilcec, R.~F., and Krasny, M. (2022).
\newblock Online learning and social norms: Evidence from a cross-cultural field experiment in a course for a cause.

\bibitem[Ciolino et~al., 2019]{Ciolino2019-ms}
Ciolino, J.~D., Diebold, A., Jensen, J.~K., Rouleau, G.~W., Koloms, K.~K., and Tandon, D. (2019).
\newblock Choosing an imbalance metric for covariate-constrained randomization in multiple-arm cluster-randomized trials.
\newblock {\em Trials}, 20(1):293.

\bibitem[Cox, 1958]{Cox1958-xg}
Cox, D.~R. (1958).
\newblock {\em Planning of experiments}.
\newblock Wiley.

\bibitem[Cox, 2009]{Cox2009-cy}
Cox, D.~R. (2009).
\newblock Randomization in the design of experiments.
\newblock {\em Int. Stat. Rev.}, 77(3):415--429.

\bibitem[Eckles et~al., 2017]{Eckles2017-sr}
Eckles, D., Karrer, B., and Ugander, J. (2017).
\newblock Design and analysis of experiments in networks: Reducing bias from interference.
\newblock {\em Journal of Causal Inference}, 5(1).

\bibitem[Fisher, 1935]{Fisher1935-mh}
Fisher, R.~A. (1935).
\newblock {\em The design of experiments}.
\newblock Oliver \& Boyd The design of experiments., Oxford, England.

\bibitem[Fisher, 1992]{Fisher1992-ed}
Fisher, R.~A. (1992).
\newblock The arrangement of field experiments.
\newblock In Kotz, S. and Johnson, N.~L., editors, {\em Breakthroughs in Statistics: Methodology and Distribution}, pages 82--91. Springer New York, New York, NY.

\bibitem[Freedman, 2008]{Freedman2008-bf}
Freedman, D.~A. (2008).
\newblock On regression adjustments to experimental data.
\newblock {\em Adv. Appl. Math.}, 40(2):180--193.

\bibitem[Greevy et~al., 2004]{Greevy2004-pd}
Greevy, R., Lu, B., Silber, J.~H., and Rosenbaum, P. (2004).
\newblock Optimal multivariate matching before randomization.
\newblock {\em Biostatistics}, 5(2):263--275.

\bibitem[Grover et~al., 2017]{Grover2017-cm}
Grover, S.~S., Ito, T.~A., and Park, B. (2017).
\newblock The effects of gender composition on women's experience in math work groups.
\newblock {\em J. Pers. Soc. Psychol.}, 112(6):877--900.

\bibitem[Hayes and Moulton, 2017]{Hayes2017-bn}
Hayes, R.~J. and Moulton, L.~H. (2017).
\newblock {\em Cluster Randomised Trials}.
\newblock Chapman and Hall/CRC, 2nd edition edition.

\bibitem[Higgins et~al., 2016]{Higgins2016-tm}
Higgins, M.~J., S{\"a}vje, F., and Sekhon, J.~S. (2016).
\newblock Improving massive experiments with threshold blocking.
\newblock {\em Proc. Natl. Acad. Sci. U. S. A.}, 113(27):7369--7376.

\bibitem[Hudgens and Halloran, 2008]{Hudgens2008-os}
Hudgens, M.~G. and Halloran, M.~E. (2008).
\newblock Toward causal inference with interference.
\newblock {\em J. Am. Stat. Assoc.}, 103(482):832--842.

\bibitem[Imai et~al., 2009]{Imai2009-fq}
Imai, K., King, G., and Nall, C. (2009).
\newblock The essential role of pair matching in {Cluster-Randomized} experiments, with application to the mexican universal health insurance evaluation.
\newblock {\em SSO Schweiz. Monatsschr. Zahnheilkd.}, 24(1):29--53.

\bibitem[Ioannidis, 2005]{Ioannidis2005-dx}
Ioannidis, J. P.~A. (2005).
\newblock Why most published research findings are false.
\newblock {\em PLoS Med.}, 2(8):e124.

\bibitem[Kasy, 2016]{Kasy2016-wo}
Kasy, M. (2016).
\newblock Why experimenters might not always want to randomize, and what they could do instead.
\newblock {\em Polit. Anal.}, 24(3):324--338.

\bibitem[Kim et~al., 2015]{Kim2015-dk}
Kim, D.~A., Hwong, A.~R., Stafford, D., Hughes, D.~A., O'Malley, A.~J., Fowler, J.~H., and Christakis, N.~A. (2015).
\newblock Social network targeting to maximise population behaviour change: a cluster randomised controlled trial.
\newblock {\em Lancet}, 386(9989):145--153.

\bibitem[Kizilcec and Cohen, 2017]{kizilcec2017eight}
Kizilcec, R.~F. and Cohen, G.~L. (2017).
\newblock Eight-minute self-regulation intervention raises educational attainment at scale in individualist but not collectivist cultures.
\newblock {\em Proceedings of the National Academy of Sciences}, 114(17):4348--4353.

\bibitem[Kizilcec et~al., 2022]{Kizilcec2022-ul}
Kizilcec, R.~F., Mimno, J.~A., and Karhan, A.~J. (2022).
\newblock Effects of framing professional development as a career growth opportunity on course completion.
\newblock In {\em Proceedings of the Ninth {ACM} Conference on Learning @ Scale}, L@S '22, pages 335--339. Association for Computing Machinery.

\bibitem[Kizilcec et~al., 2020]{kizilcec2020scaling}
Kizilcec, R.~F., Reich, J., Yeomans, M., Dann, C., Brunskill, E., Lopez, G., Turkay, S., Williams, J.~J., and Tingley, D. (2020).
\newblock Scaling up behavioral science interventions in online education.
\newblock {\em Proceedings of the National Academy of Sciences}, 117(26):14900--14905.

\bibitem[Krieger et~al., 2020]{Krieger2020-hl}
Krieger, A.~M., Azriel, D., Sklar, M., and Kapelner, A. (2020).
\newblock Improving the power of the randomization test.

\bibitem[Li et~al., 2016]{Li2016-zf}
Li, F., Lokhnygina, Y., Murray, D.~M., Heagerty, P.~J., and DeLong, E.~R. (2016).
\newblock An evaluation of constrained randomization for the design and analysis of group-randomized trials.
\newblock {\em Stat. Med.}, 35(10):1565--1579.

\bibitem[Manski, 2013]{Manski2013-ol}
Manski, C.~F. (2013).
\newblock Identification of treatment response with social interactions.
\newblock {\em Econom. J.}, 16(1):S1--S23.

\bibitem[Mitchell, 1998]{Mitchell1998-sg}
Mitchell, M. (1998).
\newblock {\em An Introduction to Genetic Algorithms}.
\newblock MIT Press.

\bibitem[Morgan and Rubin, 2012]{Morgan2012-mb}
Morgan, K.~L. and Rubin, D.~B. (2012).
\newblock Rerandomization to improve covariate balance in experiments.
\newblock {\em Ann. Stat.}, 40(2):1263--1282.

\bibitem[Moulton, 2004]{Moulton2004-sj}
Moulton, L.~H. (2004).
\newblock Covariate-based constrained randomization of group-randomized trials.
\newblock {\em Clin. Trials}, 1(3):297--305.

\bibitem[Munaf{\`o} et~al., 2017]{Munafo2017-tb}
Munaf{\`o}, M.~R., Nosek, B.~A., Bishop, D. V.~M., Button, K.~S., Chambers, C.~D., du~Sert, N.~P., Simonsohn, U., Wagenmakers, E.-J., Ware, J.~J., and Ioannidis, J. P.~A. (2017).
\newblock A manifesto for reproducible science.
\newblock {\em Nat Hum Behav}, 1(1):0021.

\bibitem[Murnane et~al., 2023]{Murnane2023-vy}
Murnane, E.~L., Glazko, Y.~S., Costa, J., Yao, R., Zhao, G., Moya, P. M.~L., and Landay, J.~A. (2023).
\newblock {Narrative-Based} visual feedback to encourage sustained physical activity: A field trial of the {WhoIsZuki} mobile health platform.
\newblock {\em Proc. ACM Interact. Mob. Wearable Ubiquitous Technol.}, 7(1):1--36.

\bibitem[Nordin and Schultzberg, 2022]{Nordin2022-jw}
Nordin, M. and Schultzberg, M. (2022).
\newblock Properties of restricted randomization with implications for experimental design.
\newblock {\em Journal of Causal Inference}, 10(1):227--245.

\bibitem[Rosenbaum, 2007]{Rosenbaum2007-xe}
Rosenbaum, P.~R. (2007).
\newblock Interference between units in randomized experiments.
\newblock {\em J. Am. Stat. Assoc.}, 102(477):191--200.

\bibitem[Rubin, 2008]{Rubin2008-va}
Rubin, D.~B. (2008).
\newblock Comment: The design and analysis of gold standard randomized experiments.
\newblock {\em J. Am. Stat. Assoc.}, 103(484):1350--1356.

\bibitem[Sacerdote, 2001]{Sacerdote2001-bg}
Sacerdote, B. (2001).
\newblock Peer effects with random assignment: Results for dartmouth roommates.
\newblock {\em Q. J. Econ.}, 116(2):681--704.

\bibitem[Sarnquist et~al., 2023]{Sarnquist2023-ho}
Sarnquist, C., Friedberg, R., Rosenman, E. T.~R., Amuyunzu-Nyamongo, M., Nyairo, G., and Baiocchi, M. (2023).
\newblock Sexual assault among young adolescents in informal settlements in nairobi, kenya: Findings from the {IMPower} and {SOS} {Cluster-Randomized} controlled trial.
\newblock {\em Prev. Sci.}

\bibitem[Tukey, 1993]{Tukey1993-no}
Tukey, J.~W. (1993).
\newblock Tightening the clinical trial.
\newblock {\em Control. Clin. Trials}, 14(4):266--285.

\bibitem[Ugander et~al., 2013]{Ugander2013-tf}
Ugander, J., Karrer, B., Backstrom, L., and Kleinberg, J. (2013).
\newblock Graph cluster randomization: network exposure to multiple universes.
\newblock In {\em Proceedings of the 19th {ACM} {SIGKDD} international conference on Knowledge discovery and data mining}, KDD, pages 329--337. Association for Computing Machinery.

\bibitem[Ugander and Yin, 2023]{Ugander2023-zk}
Ugander, J. and Yin, H. (2023).
\newblock Randomized graph cluster randomization.
\newblock {\em Journal of Causal Inference}, 11(1).

\bibitem[Valente, 2012]{Valente2012-fj}
Valente, T.~W. (2012).
\newblock Network interventions.
\newblock {\em Science}, 337(6090):49--53.

\bibitem[VanderWeele, 2015]{VanderWeele2015-iq}
VanderWeele, T. (2015).
\newblock {\em Explanation in Causal Inference : Methods for Mediation and Interaction}.
\newblock Oxford University Press, Incorporated, Oxford, UNITED STATES.

\bibitem[Walton and Yeager, 2020]{Walton2020-tr}
Walton, G.~M. and Yeager, D.~S. (2020).
\newblock Seed and soil: Psychological affordances in contexts help to explain where wise interventions succeed or fail.
\newblock {\em Curr. Dir. Psychol. Sci.}, 29(3):219--226.

\bibitem[Watson et~al., 2021]{Watson2021-ob}
Watson, S.~I., Girling, A., and Hemming, K. (2021).
\newblock Design and analysis of three-arm parallel cluster randomized trials with small numbers of clusters.
\newblock {\em Stat. Med.}, 40(5):1133--1146.

\bibitem[Xu and Kalbfleisch, 2010]{Xu2010-ha}
Xu, Z. and Kalbfleisch, J.~D. (2010).
\newblock Propensity score matching in randomized clinical trials.
\newblock {\em Biometrics}, 66(3):813--823.

\bibitem[Yeager et~al., 2019]{Yeager2019-kr}
Yeager, D.~S., Hanselman, P., Walton, G.~M., Murray, J.~S., Crosnoe, R., Muller, C., Tipton, E., Schneider, B., Hulleman, C.~S., Hinojosa, C.~P., Paunesku, D., Romero, C., Flint, K., Roberts, A., Trott, J., Iachan, R., Buontempo, J., Yang, S.~M., Carvalho, C.~M., Hahn, P.~R., Gopalan, M., Mhatre, P., Ferguson, R., Duckworth, A.~L., and Dweck, C.~S. (2019).
\newblock A national experiment reveals where a growth mindset improves achievement.
\newblock {\em Nature}, 573(7774):364--369.

\bibitem[Zahrt et~al., 2023]{Zahrt2023-jd}
Zahrt, O.~H., Evans, K., Murnane, E., Santoro, E., Baiocchi, M., Landay, J., Delp, S., and Crum, A. (2023).
\newblock Effects of wearable fitness trackers and activity adequacy mindsets on affect, behavior, and health: Longitudinal randomized controlled trial.
\newblock {\em J. Med. Internet Res.}, 25:e40529.

\bibitem[Zhou et~al., 2021]{Zhou2021-rq}
Zhou, Y., Turner, E.~L., Simmons, R.~A., and Li, F. (2021).
\newblock Constrained randomization and statistical inference for multi-arm parallel cluster randomized controlled trials.

\bibitem[de Hoop et~al., 2012]{De_Hoop2012-bm}
de Hoop, E., Teerenstra, S., van Gaal, B~.~G.~I, Moerbeek, M., Borm, G.~F. (2012).
\newblock The ``best balance'' allocation led to optimal balance in cluster-controlled trials".
\end{thebibliography}

\clearpage

\begin{thebibliography}{}

\bibitem[Tukey, 1993]{Tukey1993-no}
Tukey, J.~W. (1993).
\newblock Tightening the clinical trial.
\newblock {\em Control. Clin. Trials}, 14(4):266--285.

\bibitem[Raab and Butcher, 2001]{Raab2001-to}
Raab, G.M. and Butcher, I. (2001).
\newblock Balance in cluster randomized trials.
\newblock {\em Statistics in Medicine}, 20(3), 351--365.

\bibitem[Moulton, 2004]{Moulton2004-sj}
Moulton, L.~H. (2004).
\newblock Covariate-based constrained randomization of group-randomized trials.
\newblock {\em Clin. Trials}, 1(3):297--305.

\bibitem[Morgan and Rubin, 2012]{Morgan2012-mb}
Morgan, K.~L. and Rubin, D.~B. (2012).
\newblock Rerandomization to improve covariate balance in experiments.
\newblock {\em Ann. Stat.}, 40(2):1263--1282.

\bibitem[de Hoop et~al., 2012]{De_Hoop2012-bm}
de Hoop, E., Teerenstra, S., van Gaal, B~.~G.~I, Moerbeek, M., Borm, G.~F. (2012).
\newblock The ``best balance'' allocation led to optimal balance in cluster-controlled trials".

\bibitem[Li et~al., 2016]{Li2016-zf}
Li, F., Lokhnygina, Y., Murray, D.~M., Heagerty, P.~J., and DeLong, E.~R. (2016).
\newblock An evaluation of constrained randomization for the design and analysis of group-randomized trials.
\newblock {\em Stat. Med.}, 35(10):1565--1579.

\bibitem[Branson et~al., 2016]{Branson2016-fm}
Branson, Z., Dasgupta, T., and Rubin, D.~B. (2016).
\newblock Improving covariate balance in {2K} factorial designs via rerandomization with an application to a new york city department of education high school study.
\newblock {\em Annals of Applied Statistics}, 10(4):1958--1976.

\bibitem[Li et~al., 2017]{Li2017-yg}
Li, F., Turner, E.L., Heagerty, P.J., Murray, D.M., Vollmer, W.M. and DeLong, E.R. (2017).
\newblock An evaluation of constrained randomization for the design and analysis of group-randomized trials with binary outcomes
\newblock {\em Statistics in Medicine}, 36(24), 3791--3806.

\bibitem[Basse and Airoldi, 2018]{Basse2018-ii}
Basse, G.~W. and Airoldi, E.~M. (2018).
\newblock Model-assisted design of experiments in the presence of network-correlated outcomes.
\newblock {\em Biometrika}, 105(4):849--858.

\bibitem[Ciolino et~al., 2019]{Ciolino2019-ms}
Ciolino, J.~D., Diebold, A., Jensen, J.~K., Rouleau, G.~W., Koloms, K.~K., and Tandon, D. (2019).
\newblock Choosing an imbalance metric for covariate-constrained randomization in multiple-arm cluster-randomized trials.
\newblock {\em Trials}, 20(1):293.

\bibitem[Krieger et~al., 2020]{Krieger2020-hl}
Krieger, A.~M., Azriel, D., Sklar, M., and Kapelner, A. (2020).
\newblock Improving the power of the randomization test.

\bibitem[Watson et~al., 2021]{Watson2021-ob}
Watson, S.~I., Girling, A., and Hemming, K. (2021).
\newblock Design and analysis of three-arm parallel cluster randomized trials with small numbers of clusters.
\newblock {\em Stat. Med.}, 40(5):1133--1146.

\bibitem[Kapelner et~al., 2021]{Kapelner2021-fy}
Kapelner, A., Krieger, A.M., Sklar, M., Shalit, U., and Azriel, D. (2021).
\newblock Harmonizing Optimized Designs With Classic Randomization in Experiments
\newblock {\em The American Statistician}, 75(2), 195--206.

\bibitem[Nordin and Schultzberg, 2022]{Nordin2022-jw}
Nordin, M. and Schultzberg, M. (2022).
\newblock Properties of restricted randomization with implications for experimental design.
\newblock {\em Journal of Causal Inference}, 10(1):227--245.

\bibitem[Zhou et~al., 2021]{Zhou2021-rq}
Zhou, Y., Turner, E.~L., Simmons, R.~A., and Li, F. (2021).
\newblock Constrained randomization and statistical inference for multi-arm parallel cluster randomized controlled trials.

\bibitem[Li et~al., 2018]{Li2018-gh}
Li, X., Ding, P., and Rubin, D.~B. (2018).
\newblock Asymptotic theory of rerandomization in treatment-control experiments.
\newblock {\em Proc. Natl. Acad. Sci. U. S. A.}, 115(37):9157--9162.

\bibitem[van~der Vaart, 2012]{van-der-Vaart2012-uu}
van~der Vaart, A.~W. (2012).
\newblock {\em Cambridge series in statistical and probabilistic mathematics: Asymptotic statistics series number 3}.
\newblock Cambridge University Press, Cambridge, England.

\end{thebibliography}
\end{document}